\documentstyle[12pt,aaspp4,flushrt]{article}

\lefthead{Klessen, R.S. and Kroupa, P.}
\righthead{Dwarf Galaxies without Dark Matter}

\begin{document}

\title{ Dwarf Spheroidal Satellite Galaxies Without Dark Matter:
Results From Two Different Numerical Techniques } \author{Ralf
S. Klessen$^1$ and Pavel Kroupa$^2$\\ {\small $^1$Max-Planck-Institut
f\"{u}r Astronomie, K\"onigstuhl 17, D-69117 Heidelberg, Germany}\\
{e-mail: klessen@mpia-hd.mpg.de}\\ {\small $^2$Institut f\"{u}r
Theoretische Astrophysik, Universit\"{a}t Heidelberg,
Tiergartenstr. 15, D-69121~Heidelberg, Germany}\\ {e-mail:
pavel@ita.uni-heidelberg.de} }

\begin{abstract}
Self-consistent simulations of the dynamical evolution of a low-mass
satellite galaxy without dark matter are reported.  The orbits have
eccentricities $0.41\le e\le 0.96$ in a Galactic dark halo with a mass
of $2.85\times10^{12}\,{\rm M}_\odot$ and $4.5\times10^{11}\,{\rm
M}_\odot$.  For the simulations, a particle-mesh code with nested
sub-grids and a direct-summation N-body code running with the special
purpose hardware device {\sc Grape} are used. Initially, the satellite
is spherical with an isotropic velocity distribution, and has a mass
of $10^7\,M_\odot$. Simulations with $1.3\times10^5$ up to
$2\times10^6$ satellite particles are performed.  The calculations
proceed for many orbital periods until well after the satellite
disrupts.

In all cases the dynamical evolution converges to a remnant that
contains roughly 1~per cent of the initial satellite mass. The stable
remnant results from severe tidal shaping of the initial satellite. To
an observer from Earth these remnants look strikingly similar to the
Galactic dwarf spheroidal satellite galaxies. Their apparent
mass-to-light ratios are very large despite the fact that they contain
no dark matter. 

These computations show that a remnant without dark matter displays
larger line-of-sight velocity dispersions, $\sigma$, for more
eccentric orbits, which is a result of projection onto the
observational plane.  Assuming they are not dark matter dominated, it
follows that the Galactic dSph satellites with $\sigma>6$~km/s should
have orbital eccentricities of $e>0.5$. Some remnants have
sub-structure along the line-of-sight that may be apparent in the
morphology of the horizontal branch.

\end{abstract}


\section{Introduction}
\label{sec:intro}
At least about ten dwarf spheroidal (dSph) galaxies are known to orbit
the Milky Way at distances ranging from a few tens to a few hundred
kpc. On the sky they are barely discernible stellar density
enhancements. Some have internal substructure and appear flattened.
Their velocity dispersions are similar to those seen in globular
clusters and they have approximately the same stellar mass. However,
they are about two orders of magnitude more extended.  For spherical
systems in virial equilibrium with an isotropic velocity dispersion,
the overall mass of the system can be determined from the observed
velocity dispersion.  Comparing this `gravitational' mass to the
luminosity of the system determines the mass-to-light ratio, $M/L$ (in
the following always given in solar units ${\rm M}_{\odot}/{\rm
L}_{\odot}$). In the solar neighbourhood this ratio is $3 < M/L < 5$
(e.g. Tsujimoto et al. 1997).  Values for $M/L$ of about 10 or larger
are usually taken to imply the presence of dark matter in a stellar
system.  For the dSph satellites, $M/L$ values as large as a few
hundred are inferred, implying that these systems may be completely
dark matter dominated (for a review see Mateo 1997).

A careful compilation of the observed structural parameters and 
kinematical data for the Galactic dSph galaxies can be found in Irwin
\& Hatzidimitriou (1995). More general reviews are given by Ferguson
\& Binggeli (1994), Gallagher \& Wyse (1994), Meylan \& Prugniel
(1994), Grebel (1997) and Da Costa (1997).

There are in principle two possibilities for achieving apparent high
mass-to-light ratios without dark matter: Unresolved binary stars may
inflate the measured velocity dispersion thus increasing
$M/L$. However, this effect is not large enough for a reasonable
population of binary systems (Hargreaves et al. 1996; Olszewski et al.
1996). Assuming Newtonian gravity is valid in dSph galaxies, an
alternative may be that the assumption of virial equilibrium is
violated: the satellite galaxies might be significantly perturbed by
Galactic tides. The structural, kinematical and photometric data of
the dSph satellites show correlations that may be interpreted to be
the result of significant tidal shaping (Bellazzini, Fusi Pecci \&
Ferraro 1996). Extra-tidal stars indicate that most dSph satellites
may be losing mass (Irwin \& Hatzidimitriou 1995, Kuhn,
Smith \& Hawley 1996, Smith, Kuhn \& Hawley 1997), and Burkert (1997)
points out that, if the tidal radii derived from the Irwin \&
Hatzidimitriou (1995) profiles assuming King models are correct, then
these radii are smaller than expected from the observed large
$M/L$ values.

The ``tidal scenario'' has been studied in detail by a variety of
authors: Oh et al. (1995) modelled the evolution of dSph galaxies on
different orbits in a set of rigid spherical Galactic potentials. The
satellites are represented by $10^3$ particles and are evolved using a
softened direct N-body program over many orbital periods until
disruption. Their work allows important insights into the tidal
stability of such systems. Piatek \& Pryor (1995) concentrate on
one peri-galactic passage of a dSph galaxy in different rigid spherical
Galactic potentials. Their satellite consists of $10^4$ particles and
is modelled using a {\sc Treecode} scheme.  They find that a single
peri-galactic passage cannot perturb a satellite significantly enough
for an observer to measure a high $M/L$ ratio, reaching similar
conclusions as Oh et al. (1995). Johnston, Spergel \& Hernquist
(1995), who apply their simulations to the dynamical evolution of the
Sagittarius satellite galaxy, also arrive at similar conclusions. 

Self-consistent simulations of the long-term evolution of a low-mass
satellite galaxy on two different orbits and interacting with an
extended Galactic dark halo are presented by Kroupa (1997; hereafter
referred to as K97). The satellite consists of $3 \times 10^5$
particles and the whole system is evolved applying a particle-mesh
scheme with nested sub-grids, the aim being to study the system well
after the satellite has mostly dissolved and to `observe' its
properties as it would be seen from Earth. The satellite is projected
onto the sky and its brightness profile, line-of-sight velocity
dispersion and {\em apparent} $M/L$ ratio are determined.  These
quantities can be directly compared to the observed values for
Galactic dSph galaxies.

His main finding is that a remnant containing about 1~per cent of the
initial satellite mass remains as a long-lived and distinguishable
entity after the major disruption event. To an observer from Earth,
this remnant looks strikingly similar to a dSph galaxy.  The remnant
consists of particles that have phase-space characteristics that
reduce spreading along the orbit. That this is a possibility to be
considered had been pointed out by Kuhn (1993).  However, projection
effects are also important. An observer who's line-of-sight subtends
a small angle with the orbital path of the remnant sees an apparently
brighter satellite with internal sub-clumps and an inflated velocity
dispersion. The flattened structure that may be apparent to the
observer need not have a major axis that is oriented along the orbital
path.  The observer derives values for $(M/L)_{\rm obs}$ that are much
larger than the true mass-to-light ratio $(M/L)_{\rm true}$ of the
particles, because the object is far from virial equilibrium and has a
velocity dispersion tensor that is significantly anisotropic.

The aim of the present study is to investigate if the conclusions of
K97 can be arrived at when higher resolution simulations with more
particles are used, and when a different numerical scheme altogether
is employed. We thus hope to confirm that the high $M/L$ values of the
Galactic dSph satellites can be explained without the need for dark
matter. We furthermore suggest possible observational discriminants,
and continue the analysis of the two snapshots studied in K97.

In addition to the particle-mesh method applied in K97 and
here, we use a direct N-body integrator in connection with the special
purpose hardware device {\sc Grape}, which allows the integration of systems
with $10^5$ or more particles. It is therefore a useful tool for
studying the evolution of dSph galaxies in a Galactic dark halo.

Using two different numerical schemes enables us to determine where
the models agree, i.e. which conclusions are firm, and also where they
show deviations. This allows us to quantify uncertainties inherent to
the numerical method, but we do not aim at an in-depth discussion of
the detailed differences between the two numerical schemes.  We will
show that the general conclusions agree for both methods, and that
they can both be used equivalently to explore further regions in
parameter space.  The simulations described here are part of an
extensive study of parameter space to investigate which orbits and
which assumption for the Galactic dark halo may lead to dSph-like
objects on the sky. Detailed reports about this survey will be
presented elsewhere.

In the next section we give a short introduction to the two numerical
schemes applied here, followed by a description of the data analysis
used to evaluate the simulations. Section~\ref{sec:models} treats the
initial conditions, and in Section~\ref{sec:results} we discuss our
results.  Possible discriminants between dark-matter and tidal models
are presented in Section~\ref{sec:discriminant}.  We conclude with
Section~\ref{sec:conclusions}.

\section{Two Numerical Schemes and Data Analysis}
\label{sec:numerics}
A short description of both numerical schemes used for the simulations
is provided in Sections~\ref{subsec:GRAPE} and~\ref{subsec:SUPERBOX},
and the data reduction method is described in
Section~\ref{subsec:data}.

\subsection{Direct Integration Scheme with GRAPE}
\label{subsec:GRAPE}
{\sc Grape} is a special purpose hardware device.  Its name is an
acronym for `GRAvity PipE'. The device solves Poisson's and the force
equations for a gravitational N-body system by direct summation on a
specially designed chip, thus leading to a considerable speed-up
(Sugimoto et al. 1990, Ebisuzaki et al. 1993).  We use the currently
distributed version, {{\sc Grape}-3AF}, which contains 8 chips on one
board and therefore can compute the forces on~8 particles in parallel.
The board is connected via a standard VME interface to the host
computer, in our case a SUN Sparcstation. C and FORTRAN libraries
provide the software interface between the user's program and the
board. The computational speed of {{\sc Grape}-3AF} is approximately
$5\,$Gflops.

The force law is hardwired to be a Plummer law,

\begin{equation}
{\bf F}_i \:=\: - G \sum_{j=1}^N \frac{m_im_j({\bf r}_i-{\bf r}_j)}{(|{\bf
r}_i-{\bf r}_j|^2 + \epsilon_i^2)^{3/2}}\:.
\end{equation}

\noindent
Here $i$ is the index of the particle for which the force is
calculated and $j$ enumerates the particles which exert the force;
$\epsilon_i$ is the gravitational smoothing length of particle $i$;
$G$, $m_i$ and $m_j$ are Newton's constant and the particle masses,
respectively. We chose all particles to have the same masses and
smoothing lengths.

To increase speed, concessions in the accuracy of the force
calculations had to be made: {\sc Grape} internally works with a 20
bit fixed point number format for particle positions, with a 56 bit
fixed point number format for the forces and a 14 bit logarithmic
number format for the masses (Okumura et al. 1993). Conversion to and
from this internal number representation is handled by the interface
software. The number format limits the spatial resolution in a
simulation and constrains the force accuracy. However, for
collisionless N-body systems, the forces on a single particle 
need not be known to better than about one per cent. In
that respect, {{\sc Grape}} is comparable to the widely used {\sc Treecode}
schemes (e.g. Barnes \& Hut 1986).  

We utilise {{\sc Grape}} by implementing the direct summation approach. This
essentially involves two nested loops: an outer one over all particles
for which forces are calculated, and an inner loop for the interaction
of each of those with all other particles in the system. Therefore,
the number of operations scales as ${\cal O}(N^2)$ with the particle
number $N$. Typically, this scaling law limits the particle number to
a few thousand. However, {{\sc Grape}} substitutes the inner loop and thus a
considerable speed-up is gained. We furthermore implement variable
time-steps and interpolate the particle accelerations when no new
force calculations are needed within the required accuracy. Once the
accelerations are obtained in each time-step, the particles are
advanced using the leap-frog scheme.  The satellite galaxy is
described with $131\,072$ particles. To give an estimate for the
computational time needed: a simulation with 5500 time-steps (e.g. the
run Sat-M2 in Table~\ref{table:simulations}) typically takes three days
on a Sun Sparcstation with the {\sc Grape} board.

\subsection{SUPERBOX: A Particle-Mesh Code with Nested Sub-grids}
\label{subsec:SUPERBOX}
{\sc Superbox} is a conventional particle-mesh code (see e.g. Sellwood
1987) but allows high spatial resolution of density maxima by
employing three levels of nested grids. Each active grid has $N_{\rm
grid} = (2K)^3$ cells, where $K$ is a positive integer.  The
outermost, coarsest grid contains the local universe.  The sub-grids
of the two lower levels are positioned at the density maximum of a
galaxy and follow its motion through the coarse outer grid. In
principle any number of interacting galaxies can be treated.  The
force acting on each particle is obtained by first solving Poisson's
equation using the Fast Fourier Transform technique, and then by
numerically differentiating the potential at the position of the
particle.  The leap-frog integration scheme is used to advance 
the particles along their orbits (for a brief description of the code
see K97, a detailed account will be provided elsewhere).

For the present purpose, {\sc Superbox} is used to simulate the
interaction of two galaxies, namely of the Galactic dark halo and the
satellite galaxy.  Typically, $N_{\rm grid}=32^3$ cells per grid and
in total $1.3 \times 10^6$ particles are used. In addition, two
simulations with $N_{\rm grid}=64^3$ cells on each level and in total
$4\times10^6\,$ particles are run to test the numerical resolution of
{\sc Superbox}.  An 8000-time-step long simulation takes 5~CPU~days on
an IBM~RISC/6000~350 workstation in the first case, and in the latter
case it takes 55~CPU~days on a SUN Sparcstation~10/514.

\subsection{Data Evaluation}
\label{subsec:data}
The model satellite is analysed by reproducing terrestrial
observations of a dSph galaxy, as in  K97.

At every time-step in the simulation, the position of the density
maximum of the satellite and of its remnant is determined using the
full set of $N_{\rm sat}$ particles. Every pre-chosen number $n$ of
integration steps, a subset of $N_{\rm st}$ satellite particles is
stored on disk for the detailed analysis by the hypothetical `observer
on Earth'. In the adopted Cartesian coordinate system, this observer
is located at ${\bf R}_\odot=(0,8.5,0)$~kpc, where the origin is the
Galactic centre.

For further analysis, only those stored particles are used that have
have a distance modulus $M$ satisfying

\begin{equation}
{\rm M}_{\rm cod}-{\Delta
M\over2}\le M\le {\rm M}_{\rm cod}+{\Delta M\over2}, 
\end{equation}

\noindent
where ${\rm M}_{\rm cod}=5\,{\rm log}_{10}D_{\rm cod} - 5$ is the
distance modulus of the satellite's density maximum which lies at a
distance $D_{\rm cod}$ from the Sun, and $\Delta M$ is the magnitude
range covered by the observations.  This reduced sample is the `model
observational sample'. Throughout this paper $\Delta M = 0.8$~mag is
used, except in Section~\ref{subsec:dep}, where $\Delta M$ is varied
for a detailed study of two snapshots.

Unless stated otherwise (in Section~\ref{subsec:dep}), the
observational plane is subdivided into $k=20$ circular annuli within a
projected radial distance $r_{\rm bin}=1.5$~kpc from the density
maximum of the satellite.  These are used to evaluate the
line-of-sight velocity dispersions and the surface brightness profile.
The velocity dispersions are calculated using the iterated bi-weight
scale estimator which is the estimated dispersion about the bi-weight
mean velocity of the sample.  The bi-weight location (i.e. mean) and
scale (i.e. dispersion) estimators are described by Beers, Flynn \&
Gebhardt (1990), and are robust to outlying velocity data.

The {\it apparent} mass-to-light ratio an observer deduces is
estimated from the King-formula (see Piatek \& Pryor 1995):

\begin{equation}
\left({M\over L}\right)_{\rm obs} = {9\over 2\pi G} {\sigma_0^2
\over \mu_0 \, r_{1/2}},
\end{equation}

\noindent
where $G$ is the gravitational constant, and $r_{1/2}$ is the
half-light radius, i.e. that radius at which the projected surface
brightness density decreases by 0.75~mag/pc$^2$.  The central
line-of-sight velocity dispersion, $\sigma_0$, is calculated within
the central bin.  The central surface brightness, $\mu_0$, is
estimated by fitting an exponential surface density profile to the
`observed' radial model profile, which is obtained by counting the
number of particles in the model observational sample in the above
mentioned projected radial bins, each particle having an intrinsic
$(M/L)_{\rm true}=3$, comparable to the values derived for the solar
neighbourhood. Other values may be used to change the luminosity of
the satellite.

\section{The Models and Initial Conditions}
\label{sec:models}
The Milky Way is a highly complex stellar and gaseous system. It can
be subdivided into four major components: bulge, disk, the stellar
halo, and a non-luminous dark component required to fit the rotation
curve. The latter dominates the total mass of the system by far. We
thus simplify the problem by examining the dynamical interaction of a
satellite galaxy with this dark halo alone.  The next sub-section
details the models adopted for these two components, and
Section~\ref{subsec:ics} discusses the initial conditions for the
numerical experiments.

\subsection{Galaxy models}
\label{subsec:setup}
The dark halo of the Galaxy is taken to be an isothermal sphere with a
total mass ${\rm M}_{\rm halo}=2.85\times 10^{12}\,{\rm M}_\odot$
within 250~kpc.  This follows for a halo that is truncated at 250~kpc
and has a circular velocity of $220\:$km/s.  The crossing time of the
diameter containing 33~per cent of the halo mass, $d_{33}=137.2$~kpc,
is $t_{33}=588$~Myr. We also adopt a core radius of 5~kpc.

In the simulations with {\sc Superbox}, the dark halo is treated as a
live component consisting of $N_{\rm halo}$ particles, with $N_{\rm
halo}=1\times 10^6$ or $2 \times 10^6$.  The simulations made for a
comparison with {\sc Grape} have a halo that is cut-off at $R_{\rm
c}=40$~kpc with a total mass ${\rm M}_{\rm halo}=4.5\times
10^{11}\,{\rm M}_\odot$. In this case, the inner and middle grids have
dimensions of $30^3$~kpc$^3$ and $122^3$~kpc$^3$, respectively.  For
an inter-comparison of {\sc Superbox} simulations with a different
number of grid cells and particles, $R_{\rm c}=250$~kpc is used, in
which case the inner and middle grids have dimensions of
$50^3$~kpc$^3$ and $188^3$~kpc$^3$, respectively.  The initial
velocity dispersion is always isotropic. The isolated halo with
$R_{\rm c}=40$~kpc is allowed to relax to dynamical equilibrium by
integrating it for $9\times t_{33}$ with a time-step of 1.7~Myr.  The
halo with $R_{\rm c}=250$~kpc is integrated in isolation for $25\times
t_{33}$ with a time-step of 7~Myr.  Further details and a brief
discussion of the final slightly prolate shape of the halo with
$R_{\rm c}=250$~kpc is provided in K97.  The halo with $R_{\rm
c}=40$~kpc remains spherical after attaining dynamical
equilibrium. Both contract slightly during relaxation into
equilibrium.

In the simulations with {\sc Grape}, the dark halo of the Galaxy is a
rigid sphere with a core radius of 4~kpc, a cut-off radius of
$40\:$kpc, and a total mass ${\rm M}_{\rm halo}= 4.5\times
10^{11}\,{\rm M}_\odot$.

In all cases the satellite is initially assumed to be a Plummer sphere
with a Plummer radius $R_{\rm pl}=0.3$~kpc, a cutoff radius $R_{\rm
c}=1.5$~kpc, and a mass ${\rm M}_{\rm sat} = 10^7\,{\rm M}_\odot$.
The initial velocity dispersion is isotropic, and the crossing-time of
the diameter containing 33~per cent of the mass, $d_{33}=0.56\:$kpc,
is $84\:$Myr. The satellite model is allowed to relax to dynamical
equilibrium for typically~8 such crossing times with a time-step of
1.1~Myr ({\sc Superbox}) and 1.5~Myr ({\sc Grape}). The final,
dynamically relaxed satellite is spherical.

In the {\sc Superbox} simulation, the inner and middle grids have
dimensions $1.6^3$~kpc$^3$ and $8^3$~kpc~$^3$, respectively. The
spatial resolution is thus 50~(25)~pc per cell length within a
distance of 0.8~kpc from the satellite's density maximum, and
250~(125)~pc per cell length between 0.8~kpc and 4~kpc from the
satellite's density maximum, in the $32^3$~($64^3$) cell simulation.
There are two sets of calculations: with $N_{\rm sat} = 3 \times 10^5$
with $N_{\rm grid}=32^3$, and $N_{\rm sat} =2 \times 10^6$ with
$N_{\rm grid}=64^3$.

The calculations with {\sc Grape} use $N_{\rm sat}=131\,072$ and
$\epsilon=50$~pc (equation~1) equal for all particles.

\subsection{Initial conditions}
\label{subsec:ics}
As reasoned in the Introduction, the aim of the present paper is to
investigate, using different numerical realisations, the robustness of
the results from K97. Simulations with {\sc Superbox} are compared
with equivalent simulations running on {\sc Grape}. Also, {\sc
Superbox} simulations with in total $1.3\times10^6$ particles and
$N_{\rm grid}=32^3$ cells are compared with simulations with in total
$4\times10^6$ particles and $N_{\rm grid}=64^3$ cells. In each case,
the integration time-step is 1.1~Myr for simulations with {\sc
Superbox}, and 1.5~Myr in the lowest time-step bin for the simulations
using direct summation on {\sc Grape}.

In all simulations presented here, the satellite is initially
positioned on the $x$-axis at apo-galactic distance $R_{\rm apo}$ from
the Galactic centre, with an initial velocity vector $\bf v_{\rm 0}$
along the $y$-direction. The eccentricity of the orbit is $e= (R_{\rm
apo}-R_{\rm peri})/(R_{\rm apo}+R_{\rm peri})$, where $R_{\rm peri}$
is the peri-galactic distance.

An overview of the initial conditions for the eight simulations
described here is given in Table~\ref{table:simulations}. The first
column contains the name of the simulation, the second column ($N_{\rm
grid}$) lists the number of grid cells used with {\sc Superbox} (a $G$
indicates runs with {\sc Grape}). $N_{\rm halo}$, $N_{\rm sat}$, and
$N_{\rm st}$ are the number of halo, satellite and stored satellite
particles used in the data evaluation, respectively; $n_{\rm tot}$ is
the total number of time-steps, and every $n$ steps $N_{\rm st}$
particles are written to computer disk. Column~8 ($\Delta t$) lists
the total time interval simulated. The next two columns give the
initial centre-of-mass position, ${\bf r}_{\rm 0}$, and velocity,
${\bf v}_{\rm 0}$, vectors of the satellite in a Cartesian coordinate
system centred on the Galaxy. The last two columns list the orbital
eccentricity, $e$, and the mass of the Galactic dark halo 
(see Section~\ref{subsec:setup}).

\section{Results}
\label{sec:results}
Equivalent simulations with {\sc Superbox} and {\sc Grape} are compared
in Section~\ref{subsec:SvsG}. The dependence of the results obtained with {\sc Superbox} on the number of grid cells and particle number is discussed in
Section~\ref{subsec:cellnumber}. 

\subsection{SUPERBOX versus GRAPE}
\label{subsec:SvsG}
The initial conditions for the two pairs of {\sc Superbox} -- {\sc
Grape} simulations are listed in the top four lines of
Table~\ref{table:simulations}. The evolution of the satellite galaxy
on an orbit with eccentricity $e=0.71$ (simulations RS1-109 and
Sat-M1) and on an orbit with $e=0.46$ (RS1-113 and Sat-M2) is compared
using the two different numerical schemes. In both cases the
apo-galactic distance is $R_{\rm apo}=60$~kpc. 

Three snapshots of the satellite in simulation Sat-M1 are shown in
Fig.~\ref{fig:pos-sat-M1}. At each particular time, the satellite is
plotted as seen from outside the Galaxy (the solid line traces its
density maximum). Enclosed in the circle on the left is a magnification
of the central region of the satellite and its remnant, respectively.
The upper panel shows the satellite shortly after the start of the
calculation.  The middle panel shows the dwarf galaxy shortly after
its first apo-galacticon. Considerable tidal tails have developed, and
there is a well bound core. The bottom panel shows the galaxy shortly
after its third apo-galactic passage. The satellite has
disrupted. However, there still exists a measurable density
enhancement, the {\it remnant}, which might be identified as a dSph
satellite galaxy. This behaviour is found in all simulations studied
here and in K97.

In Fig.~\ref{fig:orbit1} we show the path of the satellite in
simulations~RS1-109 and Sat-M1 looking perpendicular on to the orbital
plane.  The satellite disrupts after the second peri-galactic passage.
Similarly, Fig.~\ref{fig:orbit2} depicts the orbit in
simulations~RS1-113 and~Sat-M2 for the first four peri-galactic
passages.

During passage through peri-galacticon the satellite is heated and
particles escape. An insightful discussion of the processes involved
is presented in Section~4 in Piatek \& Pryor (1995).  Plotting the
Lagrange radii as a function of time conveniently summarises the
overall evolution of the structure of the satellite.  The effects of
the periodic passages through peri-galacticon on the mass budget of
the satellite are shown in Fig.~\ref{fig:lagrange1} for the eccentric
orbit and in Fig.~\ref{fig:lagrange2} for the orbit with $e=0.46$.
Tidal shocks expel the outer regions of the satellites in both cases
and excite damped oscillations in those mass shells that remain bound.
High values of $(M/L)_{\rm obs}$ result only after the satellite is
completely disrupted and has reached the `remnant' phase.  This is
similar to the simulation discussed by Kuhn \& Miller (1989, see their
Fig.~2).  In their simulation, which is a simplified treatment of a
satellite on a circular orbit in a constant tidal field, the observed
mass-to-light ratio exceeds the true value by more than a factor of
five only during the disruption phase at the end.  The essential
difference is that we find long-lived remnants with significantly
inflated $(M/L)_{\rm obs}$ after the disruption event.

Comparing both numerical schemes, the evolution of the satellites are
very similar: For the eccentric orbit, the induced oscillations of the
mass shells are evident in both the {\sc Superbox} and the {\sc Grape}
simulations, and both satellites loose more than 90~per cent of the
initial mass at `disruption' time $t\approx1.3$~Gyr. The tidal forces
are milder for the less eccentric orbit ($e=0.46$), and in the {\sc
Superbox} simulation~RS1-113, the satellite disrupts at $t=3.6$~Gyr,
as is evident in Fig.~\ref{fig:lagrange2}. Disruption occurs one
orbital time (i.e. about 1.2~Gyr) later in the {\sc Grape}
simulation. However, both simulations lead to the same overall
evolution of the satellite.  A difference in disruption time between
the two simulations is to be expected because the Galactic dark halo
is taken into account in very different ways (live and self-consistent
in simulation~RS1-113, and rigid in simulation~Sat-M2) leading to
differences in the tidal forces that accumulate.

Applying the data reduction described in Section~\ref{subsec:data}, the
measured mass-to-light ratio $(M/L)_{\rm obs}$ for the satellite
remnant is very large, despite the fact that the true mass-to-light
ratio was chosen to agree with the value obtained in the solar
neighbourhood.  For both sets of simulations with {\sc Superbox} and
{\sc Grape}, $(M/L)_{\rm obs}$ is plotted in Figs.~\ref{fig:ML1}
and~\ref{fig:ML2}. These figures also show the evolution of the
central surface brightness, $\mu_0$, and of the line-of-sight velocity
dispersion, $\sigma_{r_{1/2}}$, evaluated within $r_{1/2}$, which is
largely indistinguishable from $\sigma_0$ (see
Section~\ref{subsec:ecc-orb}).

The comparison of the {\sc Superbox} simulation RS1-109 with the {\sc
Grape} simulation Sat-M1 shows excellent agreement. The apparent
mass-to-light ratio is not inflated prior to disruption despite the
forced oscillations of the Lagrange radii apparent in
Figs.~\ref{fig:lagrange1} and~\ref{fig:lagrange2}.  After disruption,
the remnant stabilises with $\mu_0\approx 10^{4.5}\,{\rm L}_\odot/{\rm
kpc}^2$.  The velocity dispersion which the observer measures for the
largest fraction of the time after satellite disruption has a value
in the range $\sigma_{r_{1/2}}\approx10-30$~km/s, and $(M/L)_{\rm
obs}$ ranges from a few hundred to a few thousand. Similarly, the
satellite on the initially less eccentric orbit ($e=0.46$) behaves
alike in the {\sc Superbox} (RS1-113) and {\sc Grape} (Sat-M2)
simulations.  With both numerical techniques, the remnant stabilises
with $\mu_0\approx 10^{4.5}\,{\rm L}_\odot/{\rm kpc}^2$,
$\sigma_{r_{1/2}}\approx2.5-10$~km/s and $(M/L)_{\rm app}\approx
100-1000$.  This remnant phase is arrived at about 1.2~Gyr later in
the {\sc Grape} simulation, owing to the earlier disruption time in
simulation RS1-113, as discussed above.

One cautionary remark is necessary at this point: The ``observed''
central surface brightness of the remnants in our simulations is, with
$\mu_0\approx 10^{4.5}\,{\rm L}_\odot/{\rm kpc}^2$, relatively low
when compared to the Galactic dSph satellites. These have central
surface luminosities ranging from $\mu_0\approx 7\times 10^5\,{\rm
L}_\odot/{\rm kpc}^2$ for Sextans to $\mu_0\approx 3\times 10^7 \,{\rm
L}_\odot/{\rm kpc}^2$ for Leo I (Irwin \& Hatzidimitriou 1995), and
are thus at least about one order of magnitude larger than our values.

However, so far we have only scanned a very small range of initial
parameters: we have limited the present study to an initial satellite
mass of $10^7\,M_\odot$. One possible way to arrive at the observed
central surface brightness is to use a satellite galaxy with an
initial mass that is approximately one order of magnitude larger.  A
simulation of a satellite galaxy with $M_{\rm sat}=10^8\,M_\odot$ on
an orbit with $e=0.85$ but with properties otherwise identical to
those of the satellite modelled here (Section~\ref{subsec:setup}),
shows that its behavior is similar to the lower mass satellites.  Due
to its higher binding energy, it needs many more orbital periods until
it is disrupted, and thus the computational cost is severe. In the
remnant phase it has $\mu_0\approx 10^{5.5}\,{\rm L}_\odot/{\rm
kpc}^2$, which is much closer to the observed values.  Its appearence
on the sky closely resembles a dSph galaxy.  Thus, if the size of the
satellite is scaled up to have the same (relative) binding energy, the
evolution is expected to be almost identical to the lower mass cases
presented here.

All values discussed here are derived adopting $(M/L)_{\rm true}=3$.
Another way to reconcile the central surface brighness of the models
presented here with the observed values, is to decrease $(M/L)_{\rm
true}$.  If we assume $(M/L)_{\rm true}=0.3$, then again $\mu_0\approx
10^{5.5}\,{\rm L}_\odot/{\rm kpc}^2$.  However, such a $(M/L)_{\rm
true}$ implies rather unusual relative numbers of bright, evolved
stars and less-luminous, unevolved stars (Dirsch \& Richtler 1995).

Finally, it is of interest to evaluate the number of particles
contributing to the central part of the remnant. To this end the
number $N$ of particles is counted in a volume with a radius of
0.8~kpc and centred on the density maximum of the remnant. While not
an exact quantification, it is a reasonable assessment of the number
of particles that are either bound energetically, or have phase-space
variables that inhibit spreading away from the remnant's density
maximum. A detailed investigation of the relative contribution of
each type of particle to $N$ awaits simulations with initially $10^7$
to $10^8$ particles per satellite galaxy.  The evolution of $N$ with
time is shown in Fig.~\ref{fig:N1} for all four runs discussed
here. As can be seen from the figure, both the {\sc Superbox} and the
{\sc Grape} simulations lead to remnants that stabilise with~0.3 to
3~per cent of the initial number of particles.  The later disruption
time of the satellite in the {\sc Grape} simulation Sat-M2 is evident,
but the outcome is the same as in the {\sc Superbox} run.

\subsection{Different number of cells and particles} 
\label{subsec:cellnumber}
The comparison between {\sc Superbox} and {\sc Grape} simulations in
the last section shows that both yield the same conclusions concerning
the evolution and fate of a low-mass satellite galaxy without dark
matter. Small differences occur, but only in as much as the time of
disruption differs by an orbital period. The satellite on the less
eccentric orbit arrives at the remnant phase one orbital period later
in the {\sc Grape} simulation. The presence of a centre to the mesh
moving with the satellite therefore does not artificially promote the
survival of a denser core in the remnant.

In this section, {\sc Superbox} simulations with $N_{\rm grid}=32^3$
cells per grid,   $N_{\rm halo}=1\times10^6$ particles in the
Galactic dark halo, and $N_{\rm sat}=3\times10^5$ particles in the
satellite are compared with {\sc Superbox} simulations with $N_{\rm
grid}=64^3$, $N_{\rm halo}=2\times10^6$ and $N_{\rm sat}=2\times10^6$.
The structure and mass of the Galactic dark halo and of the satellite
are described in Section~\ref{subsec:setup}.  Two pairs of simulations
are compared, and the initial conditions are listed in the bottom four
lines of Table~\ref{table:simulations}. Runs RS1-1 and RS1-1L simulate
the satellite galaxy on an extremely eccentric orbit ($e=0.96$), the path of
which is shown in Fig.~\ref{fig:orbit3}.  Runs RS1-24 and RS1-24L are
simulations with the satellite galaxy on an orbit with $e=0.41$. Its
trajectory is shown in Fig.~\ref{fig:orbit4}.

As is evident from Figs.~\ref{fig:lagrange3} and~\ref{fig:lagrange4},
the overall evolution of the satellite is very similar in all four
simulations. As in the {\sc Superbox} and
{\sc Grape} simulations discussed in the last section, the mass shells
are induced to oscillate with about the same period by the periodic
tidal field. Shedding of mass proceeds on about the same time scale in
the $32^3$ and $64^3$ cell simulations, although the final disruption
time is uncertain by about one orbital period.  On the highly
eccentric orbit, the satellite loses more than 90~per cent of its mass
at $t=1.4$~Gyr in the $32^3$ cell simulation (RS1-1). Disruption
occurs at $t=2.4$~Gyr in the $64^3$ cell simulation (RS1-1L). The
satellite on the less eccentric orbit is, however, disrupted at about
the same time in both simulations, RS1-24 and RS1-24L.

The evolution of the central surface brightness, of the line-of-sight
velocity dispersion within the half-light radius, and of the apparent
mass-to-light ratio for the four simulations are shown in
Figs.~\ref{fig:ML3} and~\ref{fig:ML4}. The same results are obtained,
namely that satellite evolution leads to a stable remnant that has
similar $\mu_0$, an inflated $\sigma_{r_{1/2}}$, and $(M/L)_{\rm obs}
\approx100$ or more.

It is evident that the highly eccentric orbit leads to a brighter
remnant with a larger $\sigma_{r_{1/2}}$ than the remnant on the less
eccentric orbit. This trend is also observed in
Section~\ref{subsec:SvsG} and results from projection effects, as
described in the Introduction.

In Fig.~\ref{fig:N2} the number of particles within the spherical
volume with a radius of 0.8~kpc is plotted as a function of time for
all four simulations. Again this agrees with
Section~\ref{subsec:SvsG}: the number of particles in the remnant
stabilises at~0.5 to 3.5~per cent of the initial number of satellite
particles.

\section{Possible discriminants}
\label{sec:discriminant}
The evidence presented so far shows that dark matter may not be
necessary to account for the structural and kinematical properties of
at least some of the Galactic dSph satellites. However, we do not have
unambiguous proof that this is so. Observational data that support the
tidal model include extra tidal stars (Irwin \& Hatzidimitriou 1995,
Kuhn et al.~1996, Smith et al.~1997) and substructure found in some of
the dSph systems.  These clues indicate that they may not be
completely dark matter dominated and bound systems, and that they may
be perturbed through tidal forces. However, such evidence is not fully
conclusive, because identifying extra-tidal stars depends to some
extend on the adoption of equilibrium density profiles such as given
by King models. 

There is also a linear correlation between the central surface
brightness and the integrated absolute magnitude for the Galactic dSph
systems: satellites with higher central surface brightness tend to be
{\em more} luminous. Bellazzini et al.~(1996) interpret this to be
evidence for tidal modification. This correlation is inverse to the
one found for elliptical galaxies and bulges of spiral galaxies:
systems with larger central surface brightness are {\em less} luminous
(see e.g.~Ferguson \& Binggeli 1994). A demonstration that tidal
modification of bound stellar systems leads to the observed linear
correlation between central surface brightness and integrated absolute
magnitude is to be found in Section~4 in K97. However, dwarf
elliptical galaxies in a number of galaxy clusters show the same
correlation (e.g.~Bothun, Caldwell \& Schombert 1989). Such a
correlation cannot therefore be viewed as convincing evidence for
tidal modification, unless dwarf elliptical galaxies are also tidally
modified. Finally, there is also a pronounced correlation between the
mean metallicity and luminosity of the Galactic dSph satellites, and
it is not yet clear how the present scenario can account for this.
The formation of the progenitors of dSph satellites in diffeent
locations of tidal arms pulled out of a parent galaxy that has a
radial metallicity gradient may be part of a possible solution.

In the following, we discuss further possible diagnostics to
discriminate between the dark-matter-dominated and tidal models.

\subsection{The preference for eccentric orbits}
\label{subsec:ecc-orb}  
Perusal of Figs.~\ref{fig:ML1}, \ref{fig:ML2}, \ref{fig:ML3}, and
\ref{fig:ML4} shows that during the remnant phase the average
line-of-sight velocity dispersion increases with orbital
eccentricity. To quantify this, the time-average central line-of-sight
velocity dispersion, $\left<\sigma_0\right>$, is computed over a
2.5~Gyr time interval, $t_{\rm a}-t_{\rm b}$, where $(M/L)_{\rm
obs}(t>t_{\rm a})>50$. The time-averaged line-of-sight velocity
dispersion within the half-light radius, $\left < \sigma_{r_{1/2}}
\right>$, is also computed over the same time-interval. The
simulations discussed here are augmented by the {\sc Superbox}
simulations RS1-4 and RS1-5 with $N_{\rm grid}=32^3$ cells and $N_{\rm
sat}=3\times10^5$ particles analysed in K97.  The orbits
discussed there have $e=0.74$ (RS1-4) and $0.60$ (RS1-5), with
$R_{\rm apo}=100$~kpc, and the Galactic dark halo consists of $N_{\rm
halo}=10^6$ particles, with $R_{\rm c}=250$~kpc and ${\rm M}_{\rm
halo}=2.85\times10^{12}\,{\rm M}_\odot$.

The result is plotted in Fig.~\ref{fig:vdisp_e}. The negligible
difference between the central velocity dispersion and the dispersion
evaluated within the half-light radius is evident. In this figure, the
largest differences result for simulations of nearly radial orbits,
where the modelled tidal forces at the Galactic centre are most
sensitively dependent on the resolution used. The {\sc Grape} and
high-resolution ($N_{\rm grid}=64^3$) {\sc Superbox} simulation give
essentially the same result, again nicely confirming independence of
the numerical technique used.

Fig.~\ref{fig:vdisp_e} shows that there is a well-defined correlation
between line-of-sight velocity dispersion and orbital eccentricity. In
addition, $\left < \sigma_{r_{1/2}} \right>$ is consistently larger
for the series of models with small apo-galactic distances $R_{\rm
apo}=60$~kpc, compared to the ones with $R_{\rm apo}=100\,$kpc.

The finding that $\left<\sigma_{r_{1/2}}\right>$ correlates with $e$
has possibly important implications. The velocity dispersions measured
in Galactic dSph satellites range from about 6~km/s to~11~km/s (Irwin
\& Hatzidimitriou 1995, Mateo 1997). The orbital eccentricities are
not constrained well enough yet to allow any conclusions regarding
dark matter to be made.  Oh et al. (1995) and Irwin \& Hatzidimitriou
(1995) estimate orbital eccentricities of the Galactic dSph
satellites, but theses values cannot be used here because they assume
satellite masses derived from the line-of-sight velocity dispersions.
The work reported here implies that, on average, remnants without dark
matter and with larger velocity dispersions ought to be on more
eccentric orbits.  Fig.~\ref{fig:vdisp_e} suggests that the Galactic
dSph satellites have orbital eccentricities $e>0.5$. Conversely, a
dSph satellite with $e<0.3$ and $\sigma_{r_{1/2}}>6$~km/s would have
to be dark matter dominated, unless such satellites have an extreme
internal velocity anisotropy with a large velocity dispersion
perpendicular to the direction of the orbital path (Kuhn 1993). This
special case, however, would appear to be difficult to produce under
presently understood galaxy formation mechanisms.

\subsection{Implications of $\Delta M$}
\label{subsec:dep}
Analysis of the remnants has been based on particles that lie within a
magnitude range $\Delta M = 0.8$ (equation~2) centred on the distance modulus
of the density maximum.  Observational samples used to derive {\it
kinematical} quantities for dSph galaxies typically rely on a set of
giant stars within some limited magnitude range.
 
Assuming the same intrinsic stellar brightness, this translates into a
distance selection. If the satellite is extended and has sub-structure
along the line of sight, as may be the case for tidally modified
remnants, its colour-magnitude diagram would exhibit a scatter that
might mimic populations with different ages or metallicities, and the
kinematical properties might depend on the magnitude range considered.
This is demonstrated in Fig.~12 of K97, where a significant increase
of the observed velocity dispersion is seen for increasing $\Delta
M=0.1,..., 3$~mag in one of the models. It is important to notice that
even for the smallest magnitude bin, the derived mass-to-light ratio
is extremely high and exceeds the true value by far.

{\it Structural} and {\it photometric} properties of Galactic~dSph
satellites rely on more inclusive stellar samples, because the
stringent constraint on the nature of the stars necessary for
kinematical studies (luminous stars with well defined spectral
features) can be relaxed. It is therefore important to quantify the
change of the structural parameter, $r_{1/2}$, and of the photometric
quantity, $\mu_0$, with $\Delta M$. If the tidal-modification theory
is to remain valid, then these quantities must not change much with
$\Delta M$, lest the observer would see such variations for different
sub-populations in the HR diagram of a dSph satellite.

Given that the two snapshots of satellites RS1-4 and RS1-5 at times
$t=6.27$~Gyr and $8.74$~Gyr, respectively, are studied in much detail
in sections~4.2 and~4.3 of K97, we extent that analysis here
to quantify the variation of $r_{1/2}$ and $\mu_0$ with $\Delta M$
(Section~\ref{subsubsec:strph}), and to investigate if the morphology
of the HR diagram may betray tidal modification
(Section~\ref{subsubsec:hb}). At $t=6.27$~Gyr, remnant RS1-4 has $M_{\rm
cod} = 19.33$~mag, corresponding to a Galactocentric distance of
$R_{\rm GC}=70.7$~kpc. At $t=8.74$~Gyr, remnant RS1-5 has $M_{\rm
cod} = 19.16$~mag, corresponding to a Galactocentric distance of
$R_{\rm GC}=65.7$~kpc.

\subsubsection{Dependence of structural and photometric quantities on
$\Delta M$}
\label{subsubsec:strph}
In Fig.~\ref{fig:delta-mag}, the half-light radius $r_{1/2}$ (filled
circles) and the central surface brightness $\mu_0$ (open circles) are
plotted as a function of $\Delta M$.  The two upper curves describe
the snapshot of remnant RS1-4, when its trajectory is almost aligned
with the observer's line-of-sight.  The observer sees the remnant and
parts of the leading and trailing tidal debris projected onto the same
small region in the sky, producing a considerable extend along the
line-of-sight, and thus a large variation of the observed velocity
dispersion with $\Delta M$.  However, from Fig.~\ref{fig:delta-mag} it
is apparent that the inferred $r_{1/2}$ and $\mu_0$ values are not
affected much by the sampling procedure: $r_{1/2}\approx 170 - 240$~pc
and $\mu_0\approx 10^{4.7} - 10^5 \,L_\odot/$kpc$^2$.  The snapshot of
remnant RS1-5 is observed at a larger angle to its orbital trajectory,
leading to no significant extend along the
line-of-sight. Consequently, $r_{1/2}$ and $\mu_0$ do not vary much
with $\Delta M$ (lower set of curves in Fig.~\ref{fig:delta-mag}).

\subsubsection{The width of the horizontal branch}
\label{subsubsec:hb}
A spread of distances leads to a broadening of the giant and
horizontal branches in the HR diagram.  Sub-clumping along the line-of
sight will lead to distinct populations that are separated vertically
in the HR diagram. These are important possible observational
discriminants, and the horizontal branch is especially well suited for
this type of investigation because it is horizontal and blue enough to
be less affected by contamination by foreground Galactic field stars, as
suggested by C. Pryor (private communication).  The significant
line-of-sight extension of remnant RS1-4 provides us with a model for
examining the effects of this on the horizontal branch. 

To this end, all particles are assumed to have the same luminosity, and
histograms of the number of particles in bins of distance modulus
centred on $M_{\rm cod}$ are constructed in different regions of the
remnant's face. This is done for remnants RS1-4 and RS1-5 after
storing model observational samples using $\Delta M=3$~mag
(equation~2).  The appearance of the remnants on the sky and the
distribution of mean velocities and velocity dispersions are shown in
Fig.~9 of K97, which also defines the rectilinear
coordinate system, ($x_{\rm obs},y_{\rm obs}$) on the observational
plane used here. It demonstrates that not one of
the two remnants is centred precisely on its density maximum (at
position $x_{\rm obs}=y_{\rm obs}=0\,$kpc), and that neither the
velocity gradient nor the isophotal shape of the remnant need be
aligned with the orbital trajectory.

Figures~\ref{fig:number-dM-RS1-4} and \ref{fig:number-dM-RS1-5} show
the above-mentioned histograms for models RS1-4 and RS1-5,
respectively, at three different positions along the velocity gradient
across the face of both remnants (upper panels).  The solid line
indicates the magnitude distribution of particles within a radius of
$0.2\,$kpc of the position of the density maximum of the remnants
($x_{\rm obs}=y_{\rm obs}=0$).  The long-dashed line denotes the
sample within a radial distance of 0.4~kpc of the position $x_{\rm
obs}=y_{\rm obs}=+0.8$~kpc, and the dash-dotted line is the sample
within a radial distance of 0.4~kpc of the position $x_{\rm
obs}=y_{\rm obs}=-0.8$~kpc. These three regions are aligned along the
line-of-sight velocity gradient observed in both remnants.  In each
figure, the lower panel samples all particles within a radius of
$1.2\,$kpc of $x_{\rm obs}=y_{\rm obs}=0$, thus including the above
three smaller regions.  Particles that are closer to the observer than
the density maximum have $\Delta M<0$.

The large projected depth of remnant RS1-4, together with its
clumpiness, produces a range of observed magnitudes, especially in the
lower panel in Fig.~\ref{fig:number-dM-RS1-4}. The three peaks are
separated by 0.25~mag and $-0.85$~mag, corresponding to distances of
4.2~kpc and $-12.6$~kpc, respectively, from the position of the
density maximum (70.7~kpc).  In a colour-magnitude diagram (CMD), the
apparent scatter might be interpreted as coming from three distinct
stellar components of different metallicities. However, the major peak
at $\Delta M=0$ is narrow and well defined, and would be prominent in
a CMD diagram.  The CMD of remnant RS1-5 would appear featureless and
very narrow (Fig.~\ref{fig:number-dM-RS1-5}).

\subsubsection{Words of caution}
\label{subsubsec:caut}
As argued in Section~\ref{subsec:ecc-orb}, the tidal model favours
eccentric orbits. For an observer on Earth, the angle between the
line-of-sight and the trajectory of the dSph satellite is likely to be
small, and the above projection phenomena become important. Therefore,
we expect the CMDs of at least some of these galaxies to exhibit some
scatter and possibly complex substructure.  However this prediction is
still preliminary and has to be taken with caution. We have presented
a detailed analysis of only two snapshots of the same initial low-mass
satellite on two slightly different orbits.  More general conclusions
will be possible once the parameter study now in progress is finished.
The present results do not exclude the possibility that all known
Galactic dSph satellites are more like remnant RS1-5 than RS1-4 with
colour-magnitude diagrams that are not affected by a line-of-sight
extension.

Spreads in stellar age or metallicity introduce scatter to the CMD.
Examples of the variation of the horizontal branch morphology in
dependence of metallicity and age can be found in Lee, Demarque \&
Zinn (1994).  The width of the theoretical horizontal branches is
typically~0.2 to 0.35~mag, as is true for globular clusters. In this
case the horizontal branch morphology could constrain the depth if the
particular dSph satellite has a significant extension along the
observer's line-of-sight. A more complex horizontal branch morphology
results if a dSph satellite had a complex star formation history. In
this case depth information will be difficult to extract.

The CMDs of some Galactic dSph satellites exhibit considerable
scatter, and this is usually interpreted, among other evidence in the
HR diagram, to be a sign for a complex star formation history (see
Grebel 1997 and Da Costa 1997 for excellent reviews).  A long-term aim
of the parameter study under way now, in which the orbits and initial
satellite masses are varied, is to identify those parameters that lead
to remnants that most closely resemble the known dSph satellites in
terms of $r_{1/2}$, $\sigma_0$, $\mu_0$ and $(M/L)_{\rm obs}$. A
detailed study of individual remnants will then include the
construction of synthetic CMDs.

\section{Conclusions}
\label{sec:conclusions}
Different numerical schemes are used to compute the evolution of a
low-mass satellite galaxy without dark matter on different orbits in
different spheroidal Galactic dark halos. The simulations are
performed with a particle-mesh code with nested sub-grids ({\sc
Superbox}) running on conventional workstations, and with a
direct-summation N-body code using the special-purpose hardware device
{\sc Grape}. For the former numerical approach, $32^3$ cells per
hierarchy with $3\times10^5$ satellite particles, as well as $64^3$
cells per hierarchy with $2\times10^6$ satellite particles are
used. In the latter numerical scheme, $131\,072$ satellite particles
are integrated.  The evolution is very similar and thus independent of
the numerical scheme employed. Also, the different number of grid
cells and particle number in the {\sc Superbox} simulations leads to
the same results, apart from small differences relating to the exact
time of satellite disruption.

The comparison shows that in all cases the satellite evolves to a
stable remnant that contains on the order of 1~per cent of the
original mass. This remnant phase is arrived at after the final
disruption event near peri-galacticon, during which the remaining~10
to 20~per cent of the initial satellite mass is thrown off. The
structural parameters and the line-of-sight kinematical properties are
similar to values observed for the Galactic dSph satellites. The
present model remnants have, with $(M/L)_{\rm true}=3$, lower central
surface luminosities than the Galactic dSph satellites. However,
larger initial satellite masses, or  reduced $(M/L)_{\rm
true}$ per particle, can reconcile this difference.  

Satellites initially on eccentric orbits lead to apparently brighter
remnants with inflated line-of-sight velocity dispersions, owing to
the observer's line-of-sight being approximately aligned with the
orbital path. In this case, particles ahead of and following the
remnant add to what the observer may make out to be a dSph galaxy.  An
observer looking along a very eccentric orbit finds a remnant with a
larger $\sigma$ than if its orbit were less eccentric.  The apparent
$M/L$ an observer deduces from the King formula (equation~3) is very
large, although the individual particles have $(M/L)_{\rm
true}\le3$. The line-of-sight velocity dispersion, $\sigma$, can thus
be quite unrelated to the true mass of the system.

The projection onto the observational plane has important implications
for deducing the dark matter content of Galactic dSph satellites, if
their orbital eccentricity were known.  The model data discussed here
show that the Galactic dSph satellites, which have
$\sigma_{r_{1/2}}\approx6-10$~km/s, should be on orbits with
eccentricities $e>0.5$. Conversely, if future observations confirm
$e<0.3$ for some of these systems then these must be dark matter
dominated, unless they have a very pronounced internal velocity
anisotropy (Kuhn 1993).  Such anisotropy would appear to be
difficult to produce, though, on a nearly-circular orbit, under
currently known galaxy formation mechanisms. 

The tidal model furthermore predicts scatter in the CMDs of at least
some dSph galaxies.  For preferentially eccentric orbits, projection
effects enhance the extent of the observed remnant along the
line-of-sight. This translates into a distribution of stellar
magnitudes and subsequently a scatter in the CMD, and contributes to
the scatter usually interpreted to be a sign of complex star
formation histories or metallicity variations. The morphology of the
horizontal branch is well suited for the study of possible projection
effects. The best candidate Galactic dSph satellites to show such an
effect are those with internal sub-clumps, which cannot be expected to
be at exactly the same distance. Thus, each sub-clump may contribute a
horizontal branch displaced vertically by a few tenths of a magnitude
relative to the horizontal branches of the other sub-clumps.

This work strengthens the evidence that at least some of the dSph
satellite galaxies may not be dark-matter dominated, confirming the
conclusions of K97. It follows that the interpretation by
Kuhn (1993), that at least some of the Galactic dSph satellite galaxies
may be systems with special phase-space characteristics that permit
long-time survival, is to be taken seriously.  Self-consistent
simulations of the type analysed here show how a tidally-shaped
remnant may be obtained through periodic modifications of the
phase-space properties of the satellite particles. 

The conclusions arrived at here are in agreement with the suggestion
that some of the {\it progenitors} of dwarf spheroidal galaxies
surrounding the Milky Way may have formed as condensations in tidal
arms of past merging events (see e.g. Lynden-Bell \& Lynden-Bell 1995,
and Grebel 1997 for a review). The formation of dwarf galaxies in
tidal tails is studied in detail by Elmegreen, Kaufman \& Thomasson
(1993) and Barnes \& Hernquist (1992).


\acknowledgements We thank A.~Burkert, D.~Pfenniger, and R.~Spurzem
for many stimulating discussions. We are especially grateful to
C.~Pryor for his insightful comments and suggestions that
significantly improved this article, and to G.~Bothun for his help.
The velocity dispersions were calculated using the {\sc Rostat}
software package kindly supplied by T.C. Beers. PK acknowledges
support through the Sonder-Forschungs-Bereich~328.


\clearpage


{
\begin{table}
{\small
\begin{minipage}[t]{20cm}
\hspace{-2.1cm}
  \begin{tabular}{*{16}{c}}
   \tableline\tableline
    Simulation & $N_{\rm grid}$ & $N_{\rm halo}$ & $N_{\rm sat}$ 
    &$N_{\rm st}$ &$n_{\rm tot}$ & $n$ &$\Delta t$ 
    &${\bf r}_{\rm 0}$  
    &${\bf v}_{\rm 0}$    &$e$ &${\rm M}_{\rm halo}$\\
    \tableline
              &                 &                &               &
    &             &              &{\scriptsize (Gyr)} 
    &{\scriptsize(kpc)}
    &{\scriptsize(km/s)}  &    &{\scriptsize ($\times10^{12}\,{\rm M}_\odot$)}\\
    \tableline\tableline
    RS1-109 & $32^3$ & $1 \times 10^6$ & $3 \times 10^5$ & $5\times 10^4$ &
    8000 & 30 & 8.8 & 
    $(\:60 \:/\: 0 \:/\:0\:)$ &  $(\:0 \:/\: 60 \:/\:0\:)$ &0.71  &0.45\\ 
    Sat-M1 & $G$ & --- & $131\,072$ & $65\,536$ & 3000 & 15 & 4.5& 
     $(\:60 \:/\: 0 \:/\:0\:)$ &  $(\:0 \:/\: 60 \:/\:0\:)$ &0.71 &0.45\\ 
    \tableline
    RS1-113 & $32^3$ & $1 \times 10^6$ & $3 \times 10^5$ & $5 \times 10^4$ &
    6000 & 30 & 6.6 &
    $(\:60 \:/\: 0 \:/\:0\:)$ &  $(\:0 \:/\: 120 \:/\:0\:)$ &0.46 &0.45\\ 
    Sat-M2 & $G$ & --- & $131\,072$ & $65\,536$ & 5500 & 15 & 8.25 & 
    $(\:60 \:/\: 0 \:/\:0\:)$ &  $(\:0 \:/\: 120 \:/\:0\:)$ &0.46 &0.45\\ 
    \tableline \tableline
    RS1-1L & $64^3$ & $2 \times 10^6$ & $2 \times 10^6$ & $10^5$ &
    4500 & 15 & 5.0 &
    $(\:100 \:/\: 0 \:/\:0\:)$ &  $(\:0 \:/\: 25 \:/\:0\:)$ &0.96 &2.85\\ 
    RS1-1  & $32^3$ & $1 \times 10^6$ & $3 \times 10^5$ & $5 \times
    10^4$ & 5000 & 15 & 5.5 &
    $(\:100 \:/\: 0 \:/\:0\:)$ &  $(\:0 \:/\: 25 \:/\:0\:)$ &0.96 &2.85\\ 
    \tableline
    RS1-24L & $64^3$ & $2 \times 10^6$ & $2 \times 10^6$ & $10^5$
    & 7500 & 40 & 8.3 &
    $(\:60 \:/\: 0 \:/\:0\:)$ &  $(\:0 \:/\: 175 \:/\:0\:)$ &0.41 &2.85\\ 
    RS1-24 & $32^3$ & $1 \times 10^6$ & $3 \times 10^5$ & $5 \times 10^4$ &
    10000 & 60 & 11 & 
    $(\:60 \:/\: 0 \:/\:0\:)$ &  $(\:0 \:/\: 175 \:/\:0\:)$ &0.41 &2.85\\ 
    \tableline\tableline
\end{tabular}
\end{minipage}
}

\caption{\label{table:simulations}
Table of initial conditions for the simulations presented here.}
\end{table}
}

\clearpage
\newpage

\begin{center}
{\footnotesize FIGURE CAPTIONS}
\end{center}

\begin{figure}[ht]
\epsscale{0.4}
\plotone{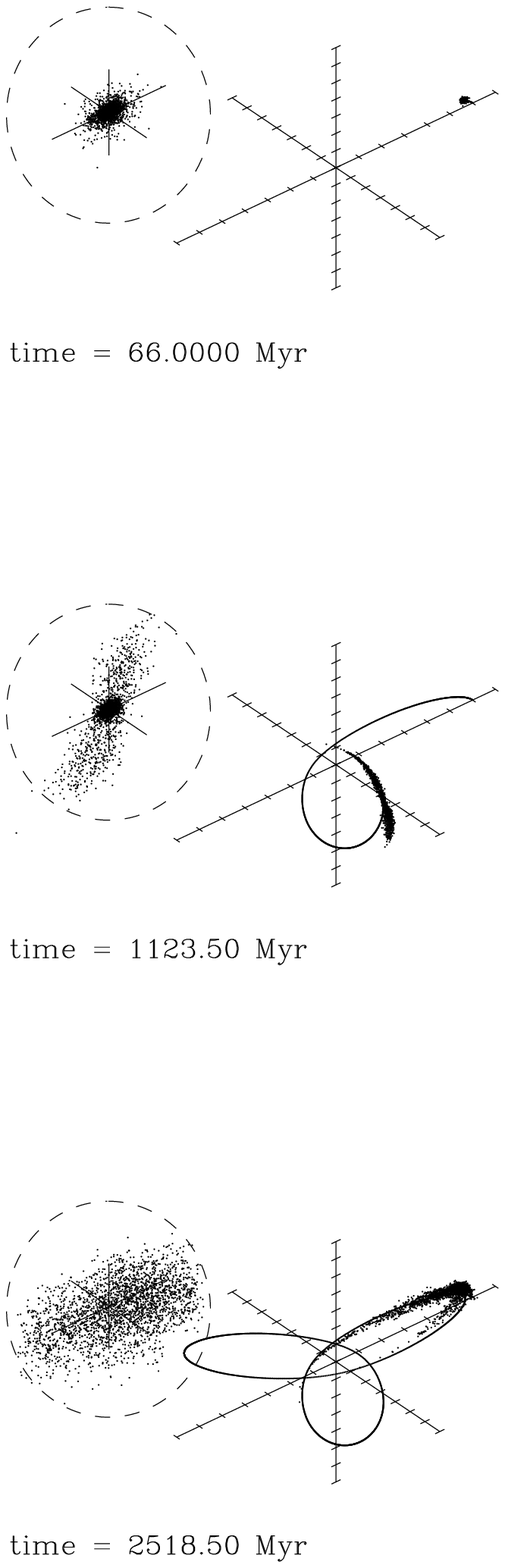}
\caption{\label{fig:pos-sat-M1}}
\end{figure}

{\normalsize F{\footnotesize IG}.}\ref{fig:pos-sat-M1} --- Snapshot of
the evolution of the satellite in {\sc Grape} simulation Sat-M1 at
three different times. The right side of each panel plots the
distribution of satellite particles at the given time in the Galactic
coordinate system. Each of the axes is 140$\:$kpc long. The solid line
depicts the trajectory of the density maximum of the satellite until
the time of the snapshot. On the left, the central part of the
satellite is shown enlarged (the total length of each axis is
5$\:$kpc).

\begin{figure}[ht]
\epsscale{0.4}
\plotone{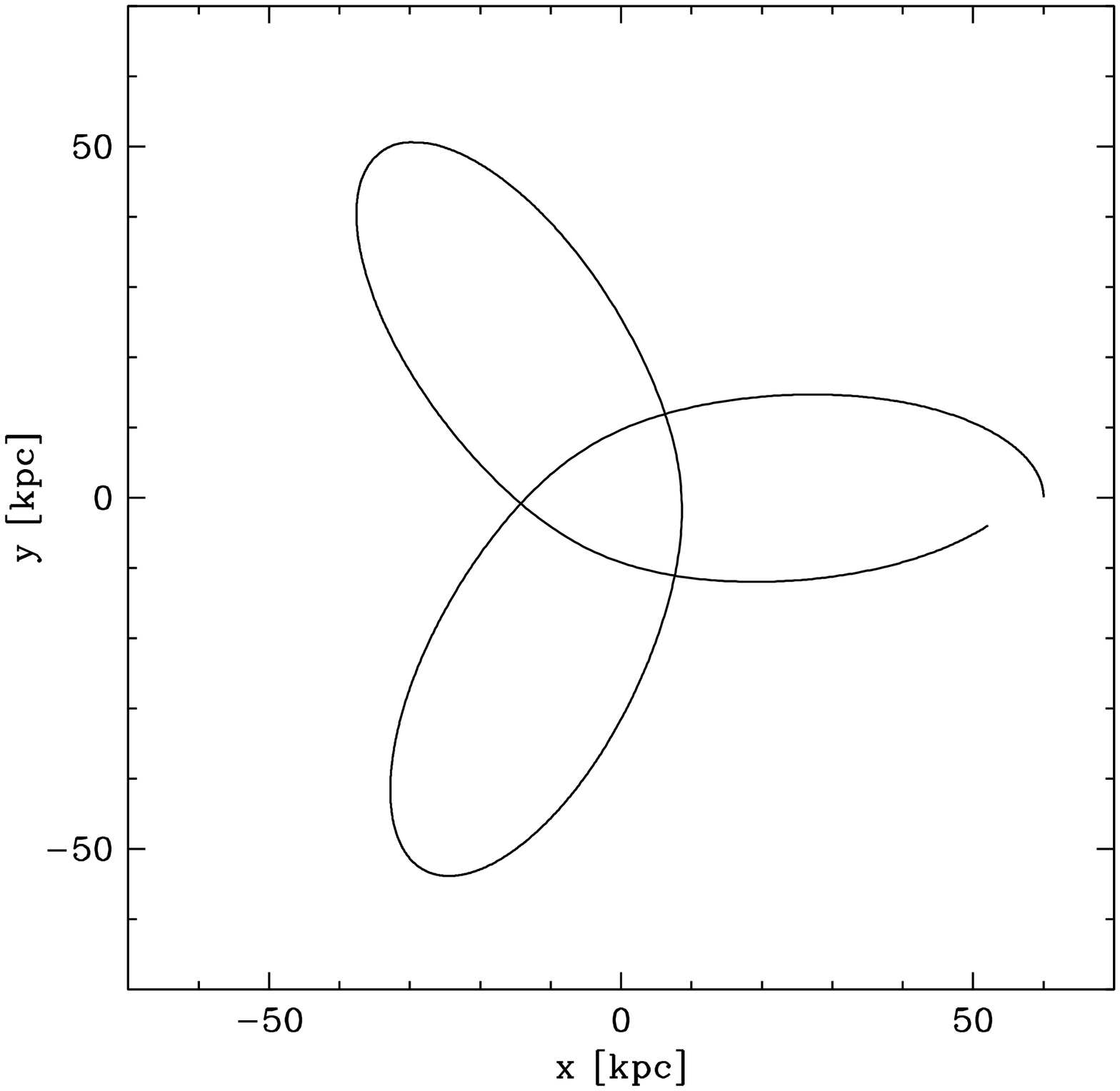}
\caption{\label{fig:orbit1}}
\end{figure}

{\normalsize F{\footnotesize IG}.}\ref{fig:orbit1} --- The orbital
path of the satellite in simulations RS1-109 and Sat-M1.

\begin{figure}[ht]
\epsscale{0.4}
\plotone{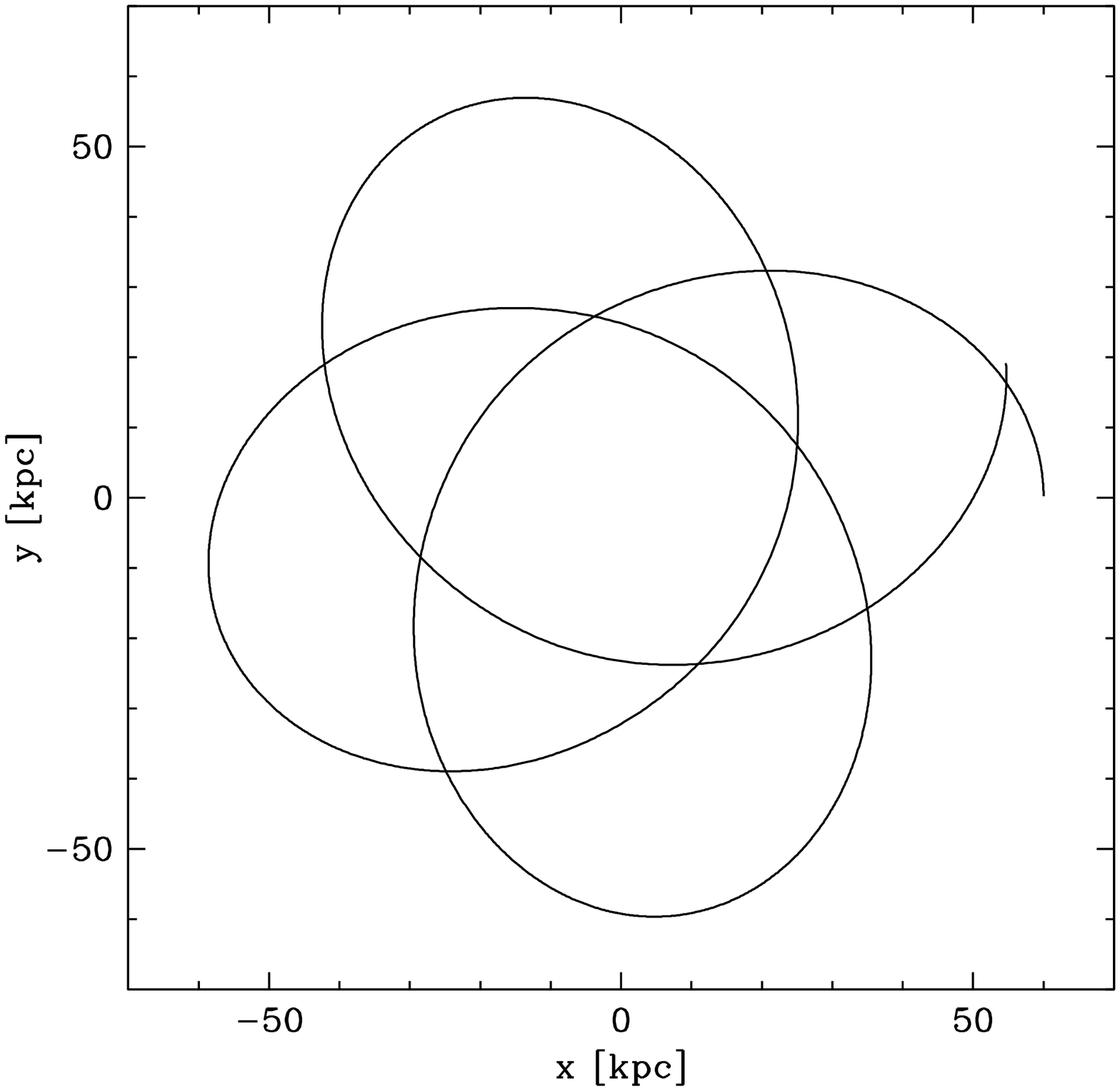}
\caption{\label{fig:orbit2}}
\end{figure}

{\normalsize F{\footnotesize IG}.}\ref{fig:orbit2} --- The orbital
path of the satellite in simulations RS1-113 and Sat-M2.

\begin{figure}[ht]
\epsscale{0.4}
\plotone{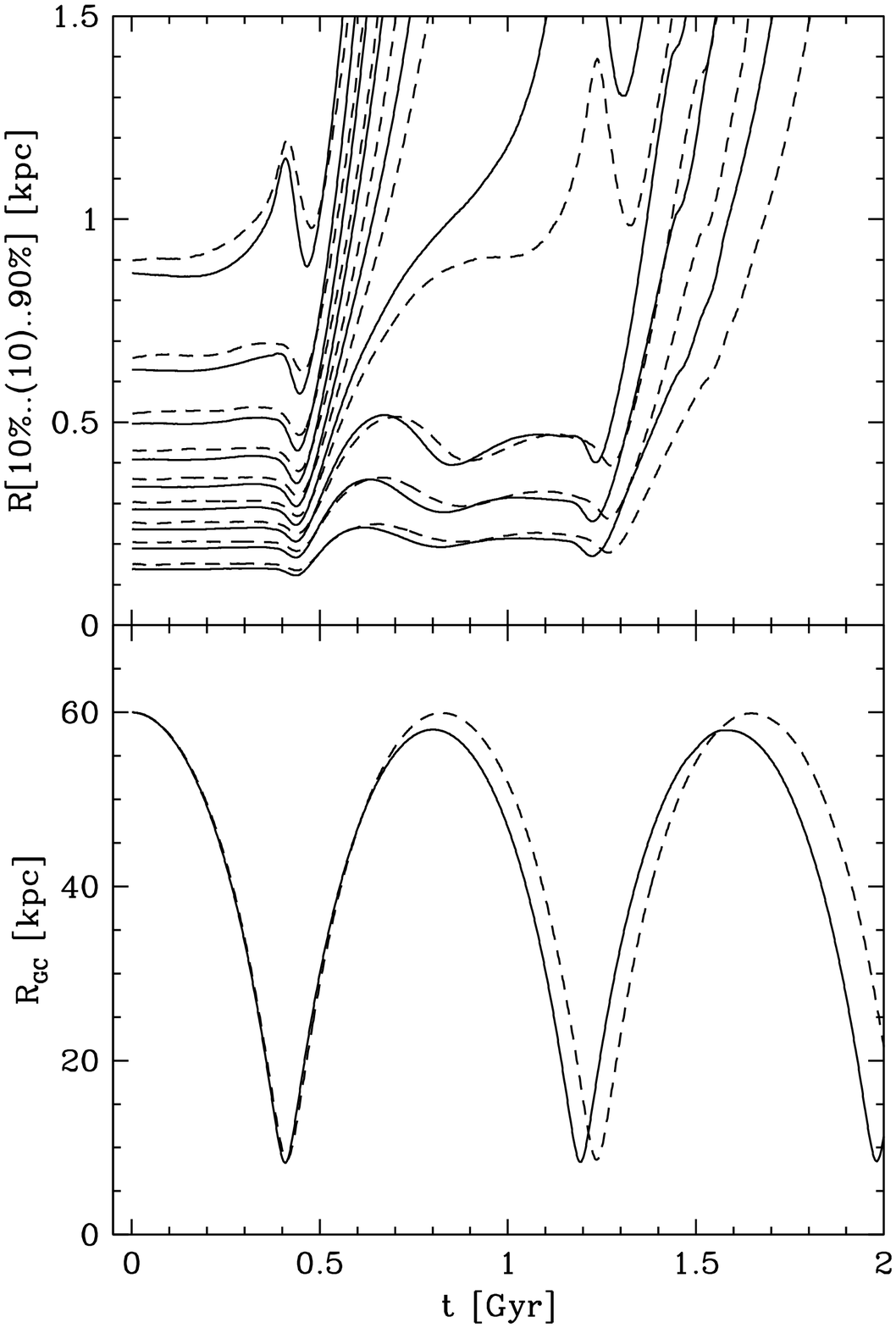}
\caption{\label{fig:lagrange1}}
\end{figure}

{\normalsize F{\footnotesize IG}.}\ref{fig:lagrange1} --- The upper
panel shows the evolution of the radii containing 10, 20, \dots,
90~per cent of the total mass of the satellite, and the lower panel
shows the Galactocentric distance as a function of time.  In both
panels the solid curve is for simulation RS1-109 ({\sc Superbox}), and
the dashed curve is for Sat-M1 ({\sc Grape}).

\begin{figure}[ht]
\epsscale{0.4}
\plotone{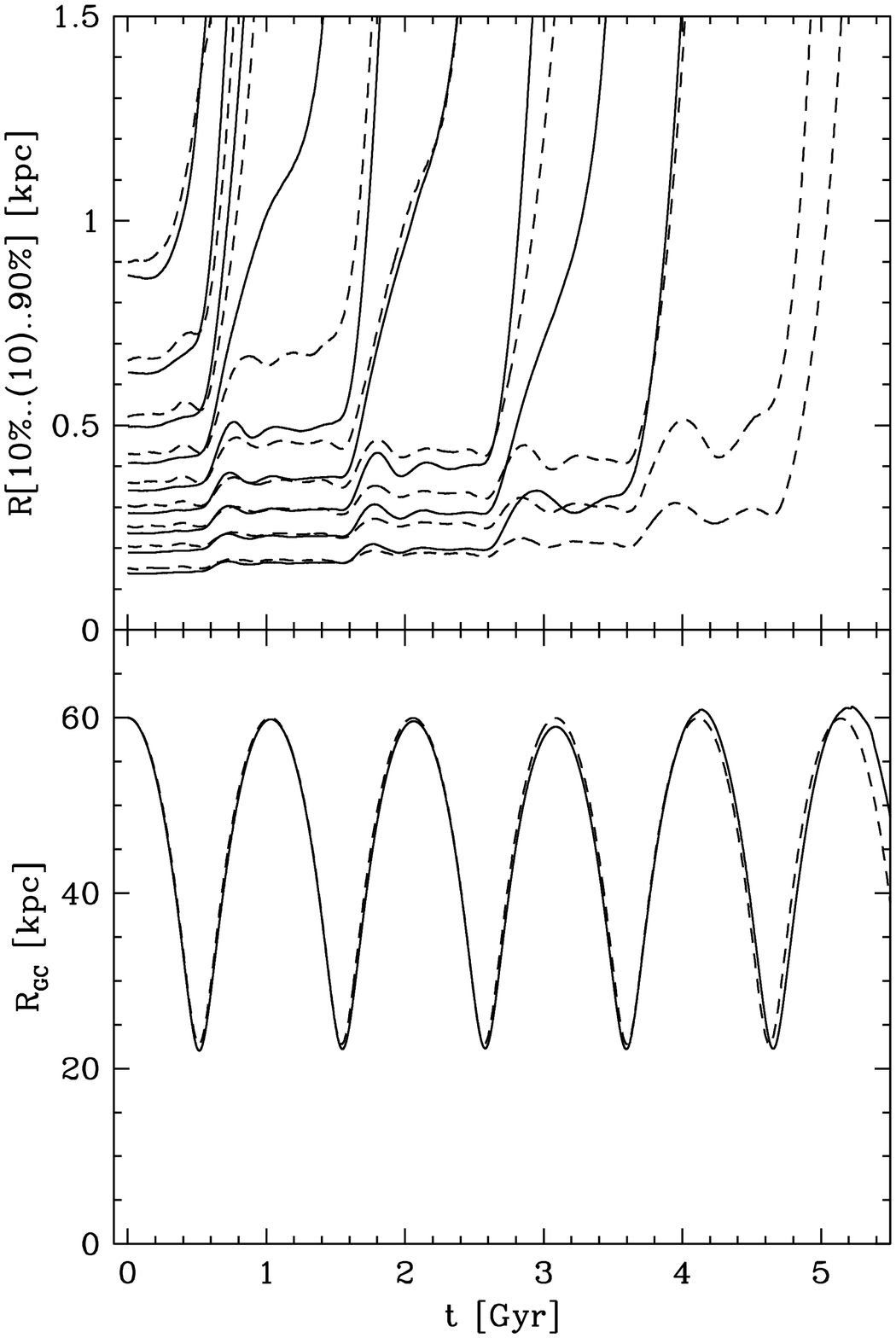}
\caption{\label{fig:lagrange2}}
\end{figure}

{\normalsize F{\footnotesize IG}.}\ref{fig:lagrange2} --- As
Fig.~\ref{fig:lagrange1}, but for {\sc Superbox} simulation RS1-113
(solid curve) and {\sc Grape} simulation Sat-M2 (dashed line).

\begin{figure}[ht]
\epsscale{0.4}
\plotone{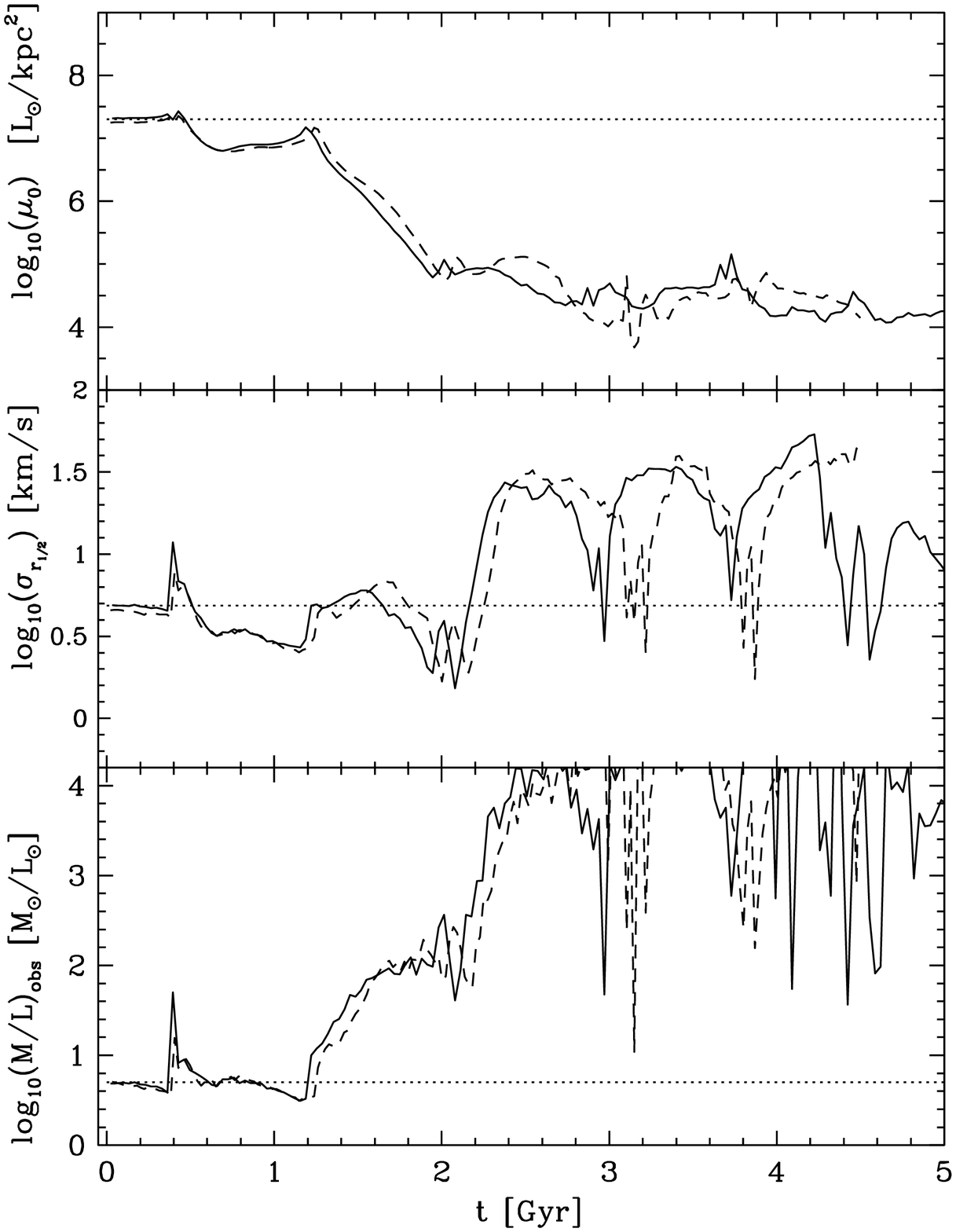}
\caption{\label{fig:ML1}}
\end{figure}

{\normalsize F{\footnotesize IG}.}\ref{fig:ML1} --- Upper panel:
evolution of the central surface brightness. Central panel: evolution
of the line-of-sight velocity dispersion within the half-light radius,
$r_{1/2}$. Bottom panel: evolution of the apparent mass-to-light ratio
evaluated using equation~3. In all panels, the solid line is {\sc
Superbox} simulation RS1-109, and the dashed line is {\sc Grape}
simulation Sat-M1. 

\begin{figure}[ht]
\epsscale{0.4}
\plotone{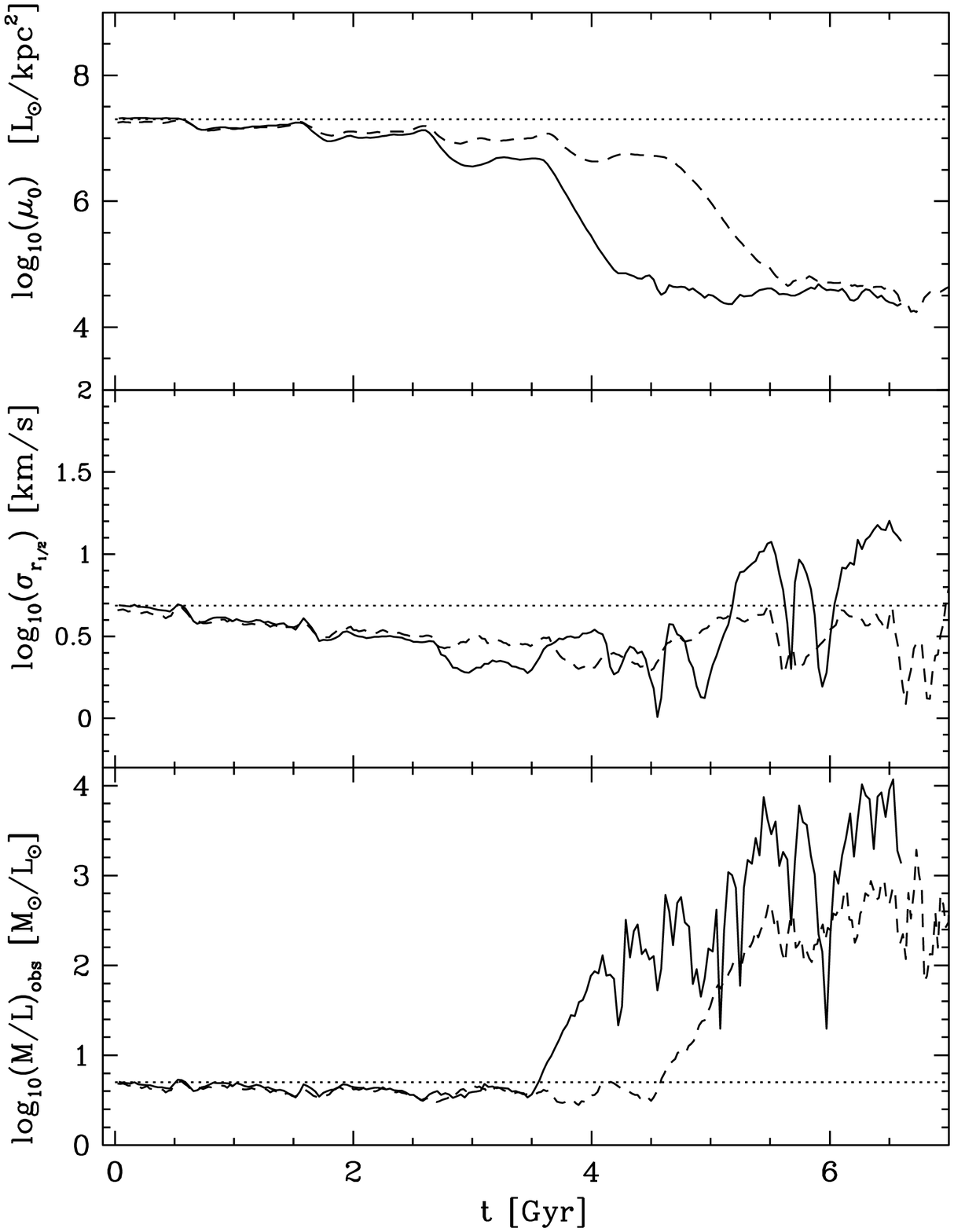}
\caption{\label{fig:ML2}}
\end{figure}

{\normalsize F{\footnotesize IG}.}\ref{fig:ML2} --- As
Fig.~\ref{fig:ML1}, but for {\sc Superbox} simulation RS1-113 (solid
line), and {\sc Grape} simulation Sat-M1 (dashed line).

\begin{figure}[ht]
\epsscale{0.4}
\plotone{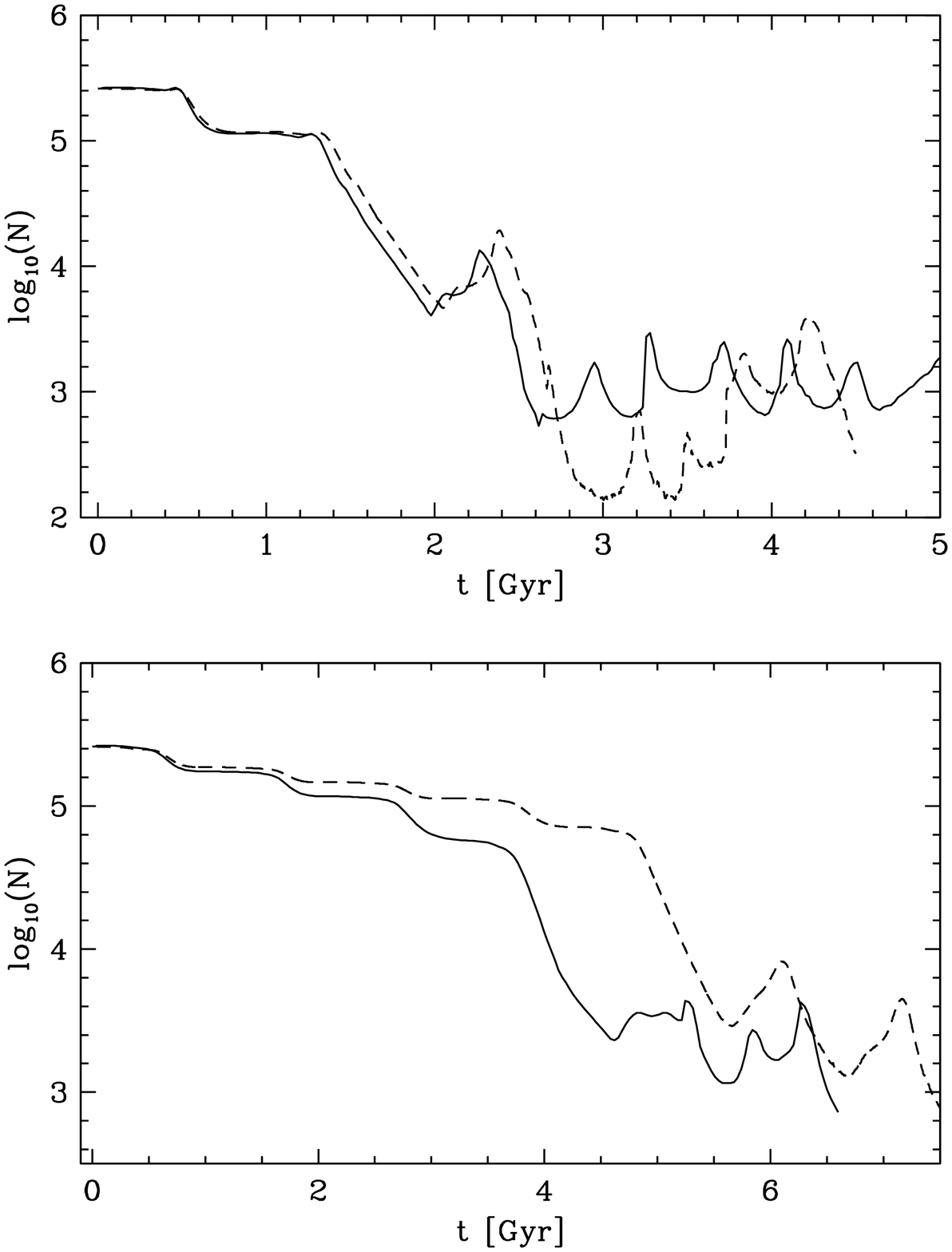}
\caption{\label{fig:N1}}
\end{figure}

{\normalsize F{\footnotesize IG}.}\ref{fig:N1} --- The number of
particles $N(t)$ in the volume with radius 0.8~kpc centred on the
density maximum of the satellite. Upper panel: the solid line is {\sc
Superbox} simulation RS1-109, and the dashed line is {\sc Grape}
simulation Sat-M1. Lower panel: the solid line is {\sc Superbox}
simulation RS1-113, and the dashed line is {\sc Grape} simulation
Sat-M2. In both panels, the number of {\sc Grape} particles is scaled
up by the factor $(3\times10^5)/131072$, where the nominator and
denominator are the initial number of particles in the {\sc Superbox}
and {\sc Grape} simulations, respectively.

\begin{figure}[ht]
\epsscale{0.4}
\plotone{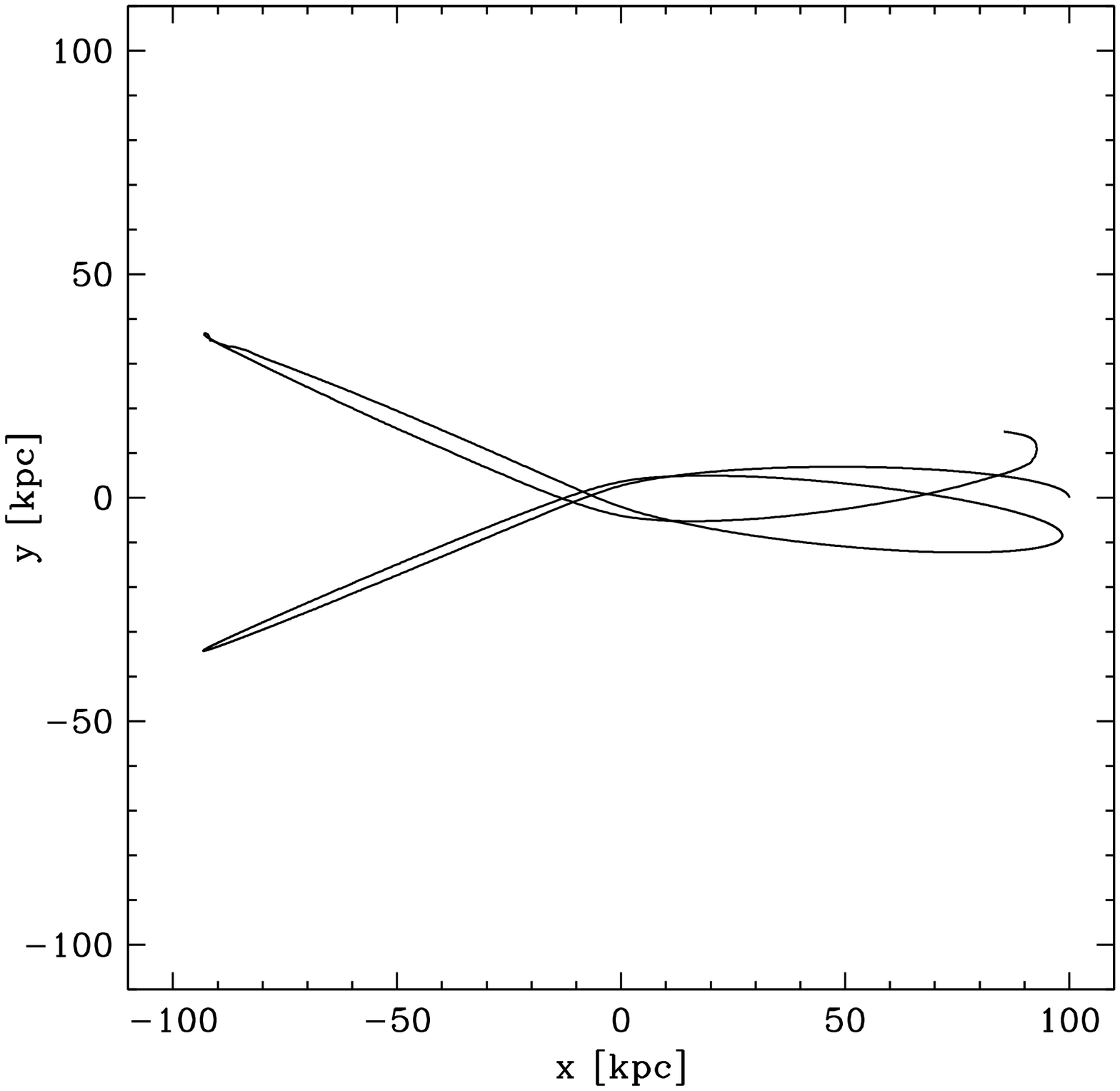}
\caption{\label{fig:orbit3}}
\end{figure}

{\normalsize F{\footnotesize IG}.}\ref{fig:orbit3} --- The orbital
path of the satellite in the $32^3$~cell simulation RS1-1 and the
$64^3$~cell simulation RS1-1L.  Owing to the slight prolate form of
the live Galactic dark halo, the orbital plane flips near
apo-galacticon. This renders the projection of the orbital plane onto
the $x$-$y$ plane somewhat irregular.

\begin{figure}[ht]
\epsscale{0.4}
\plotone{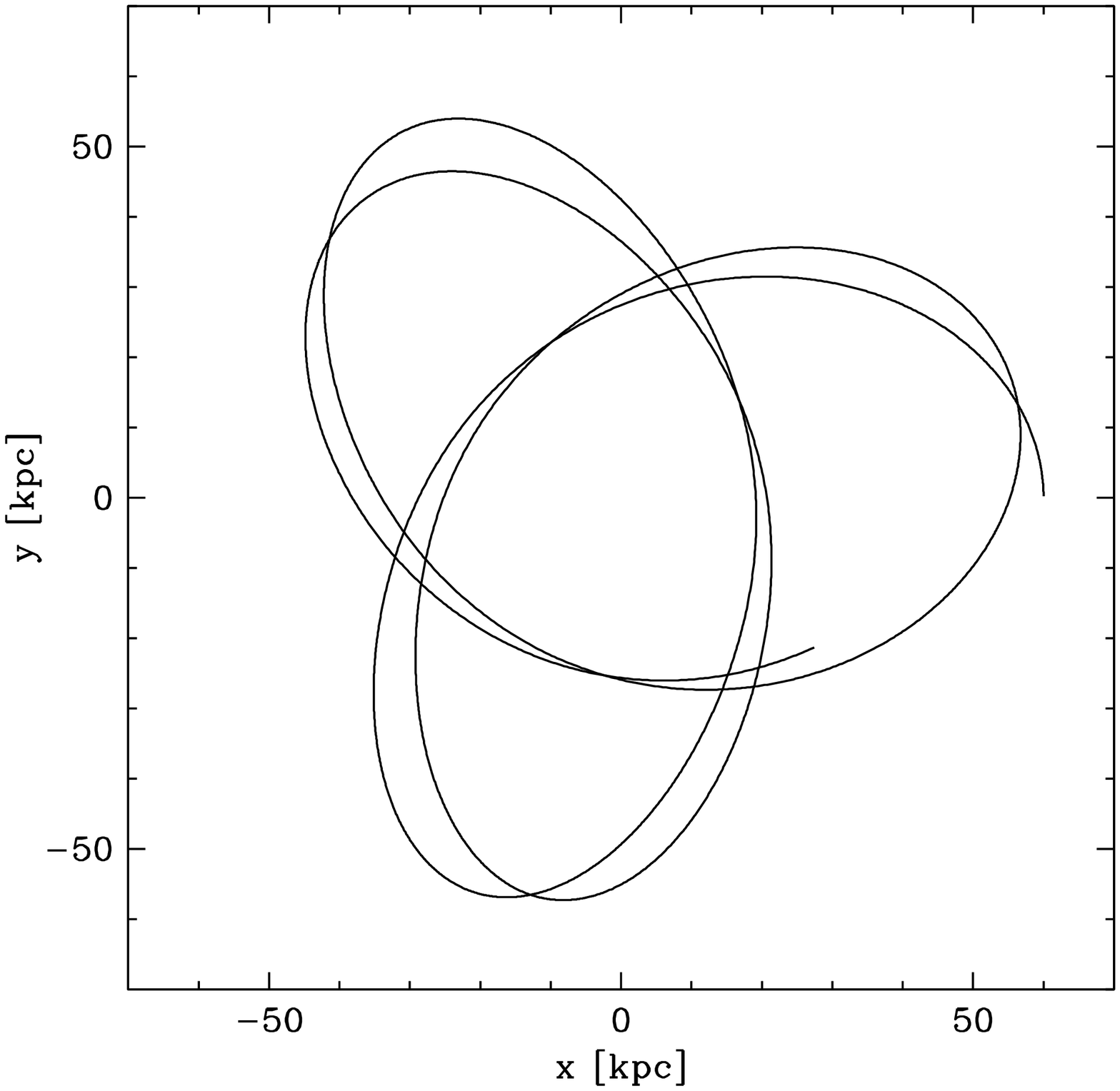}
\caption{\label{fig:orbit4}}
\end{figure}

{\normalsize F{\footnotesize IG}.}\ref{fig:orbit4} --- The orbital
path of the satellite in $32^3$~cell simulation RS1-24 and $64^3$~cell
simulation RS1-24L.

\begin{figure}[ht]
\epsscale{0.4}
\plotone{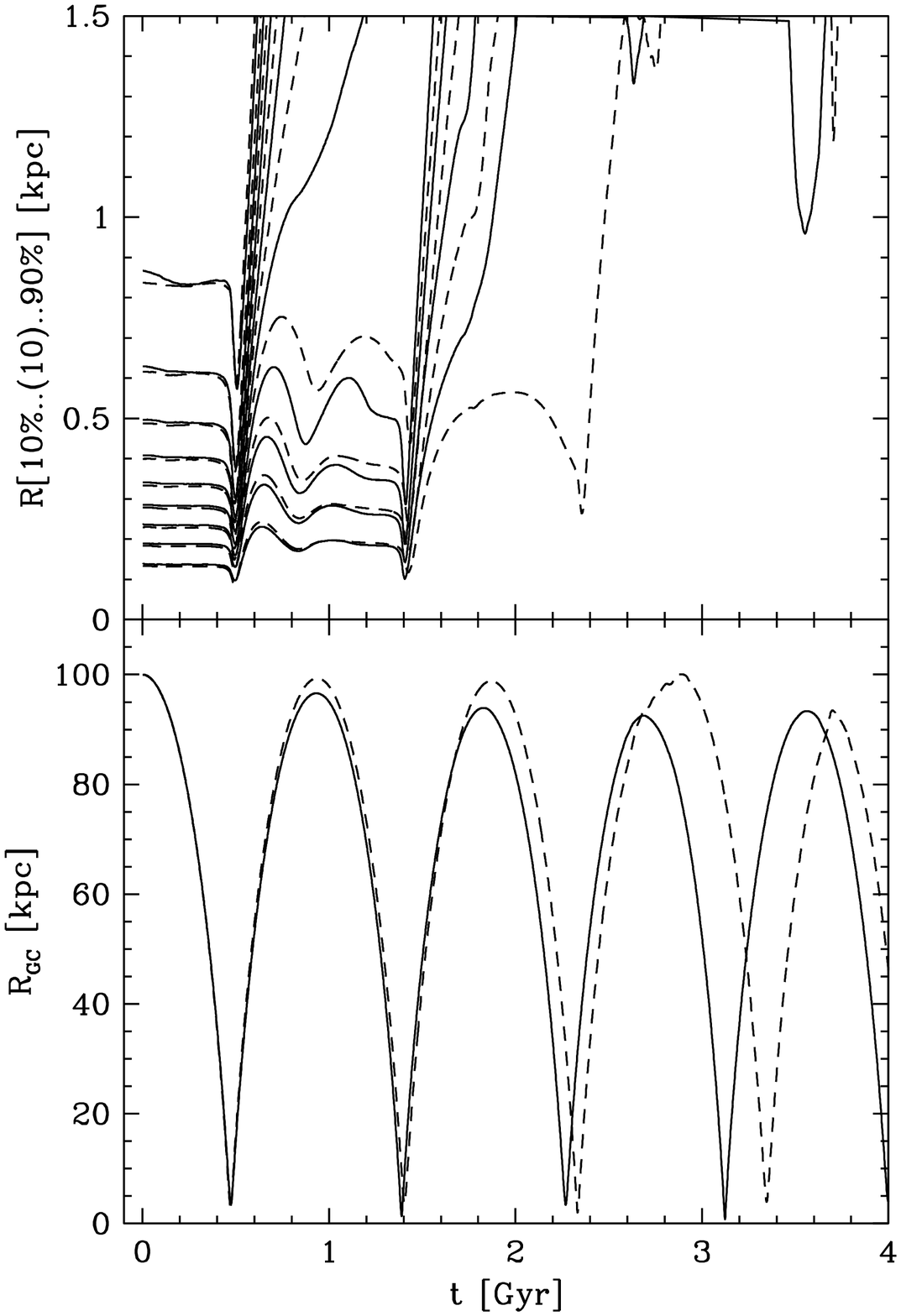}
\caption{\label{fig:lagrange3}}
\end{figure}

{\normalsize F{\footnotesize IG}.}\ref{fig:lagrange3} --- As
Fig.~\ref{fig:lagrange1}, but for $32^3$~cell simulation RS1-1 (solid
curve) and $64^3$~cell simulation RS1-1L (dashed lines).

\begin{figure}[ht]
\epsscale{0.4}
\plotone{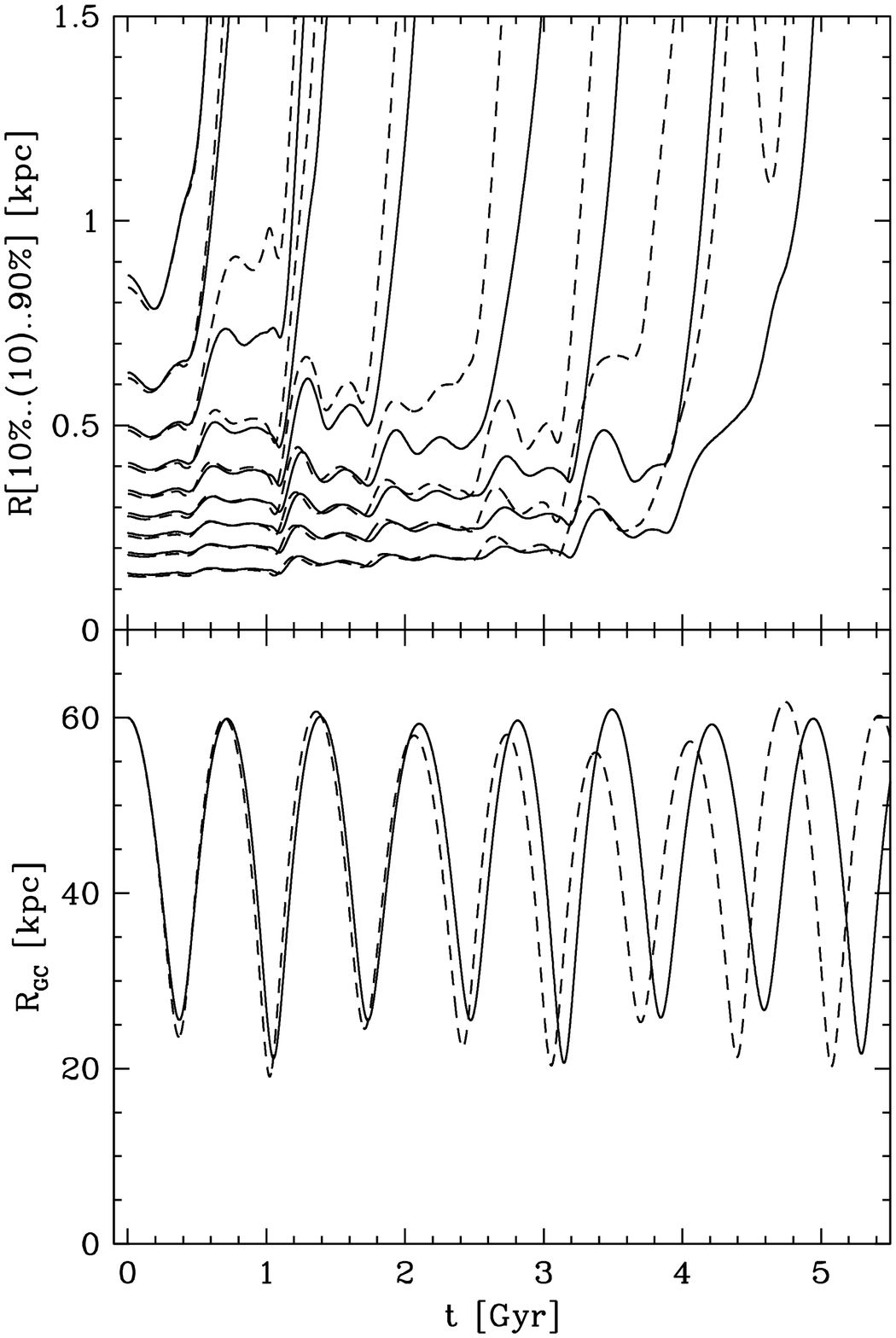}
\caption{\label{fig:lagrange4}}
\end{figure}

{\normalsize F{\footnotesize IG}.}\ref{fig:lagrange4} --- As
Fig.~\ref{fig:lagrange1}, but for $32^3$~cell simulation RS1-24 (solid
curve) and $64^3$~cell simulation RS1-24L (dashed lines).

\begin{figure}[ht]
\epsscale{0.4}
\plotone{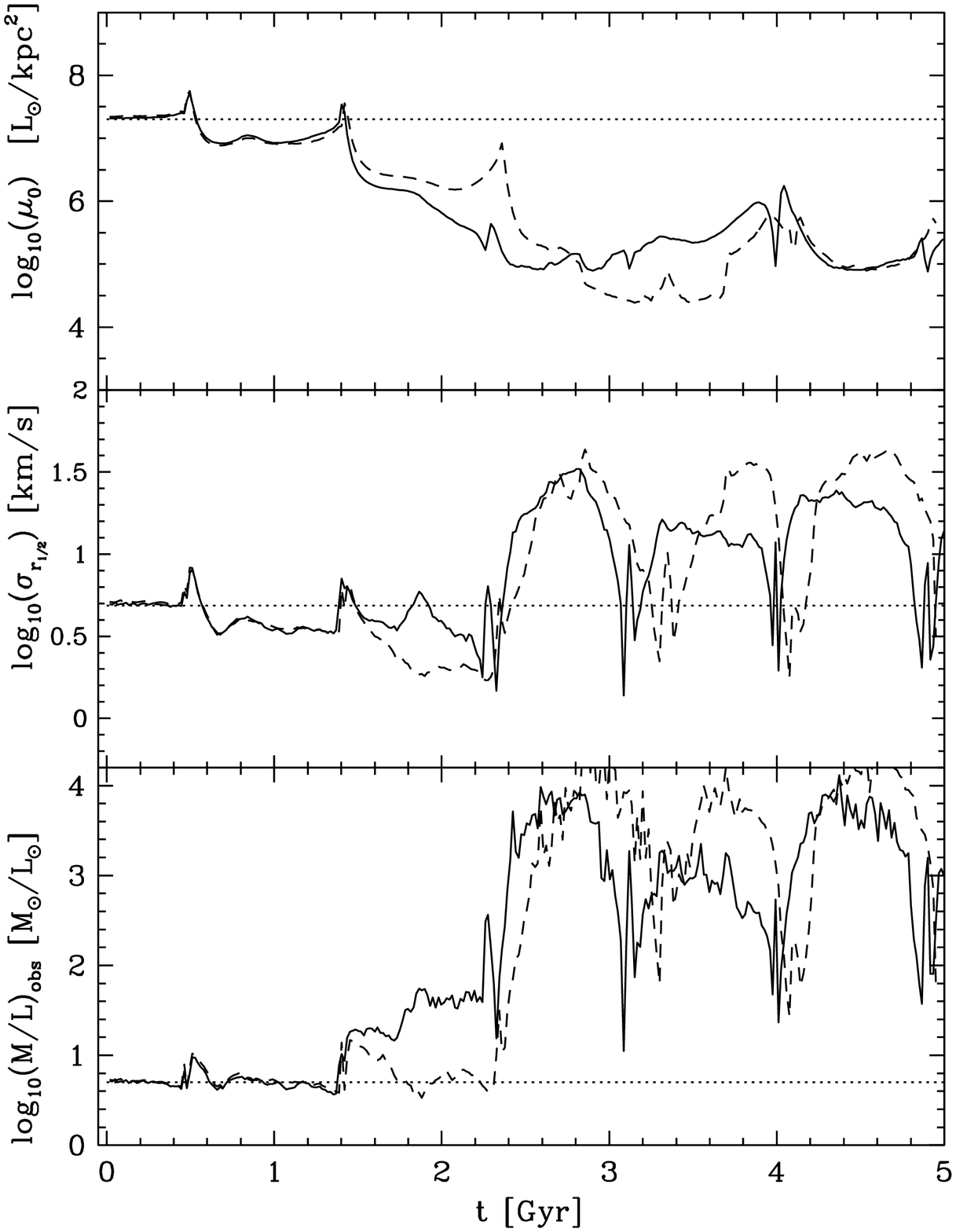}
\caption{\label{fig:ML3}}
\end{figure}

{\normalsize F{\footnotesize IG}.}\ref{fig:ML3} --- As
Fig.~\ref{fig:ML1}, but for $32^3$~cell  simulation RS1-1 (solid
line) and $64^3$~cell simulation RS1-1L (dashed line).

\begin{figure}[ht]
\epsscale{0.4}
\plotone{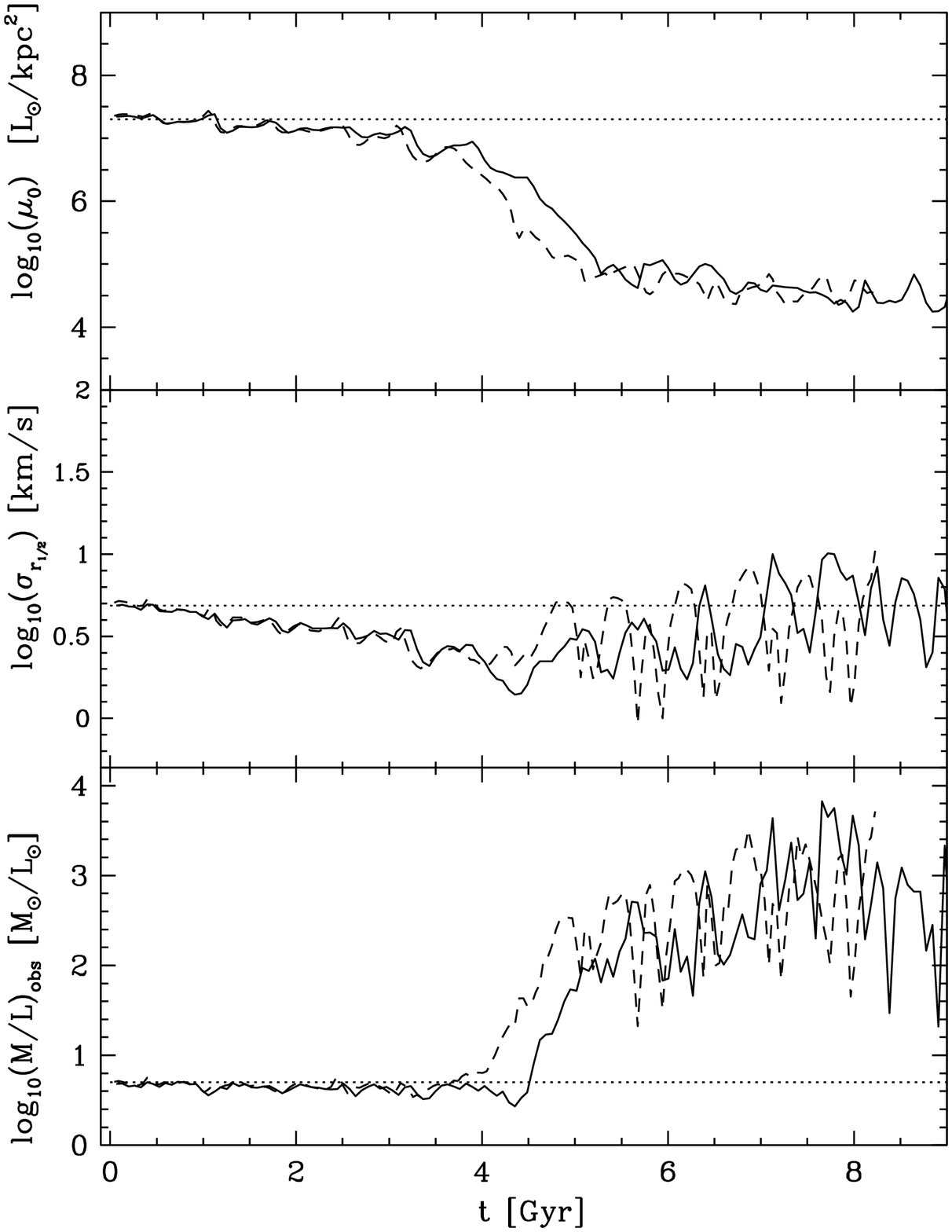}
\caption{\label{fig:ML4}}
\end{figure}

{\normalsize F{\footnotesize IG}.}\ref{fig:ML4} --- As
Fig.~\ref{fig:ML1}, but for $32^3$~cell simulation RS1-24 (solid
line) and $64^3$~cell simulation RS1-24L (dashed line).

\begin{figure}[ht]
\epsscale{0.4}
\plotone{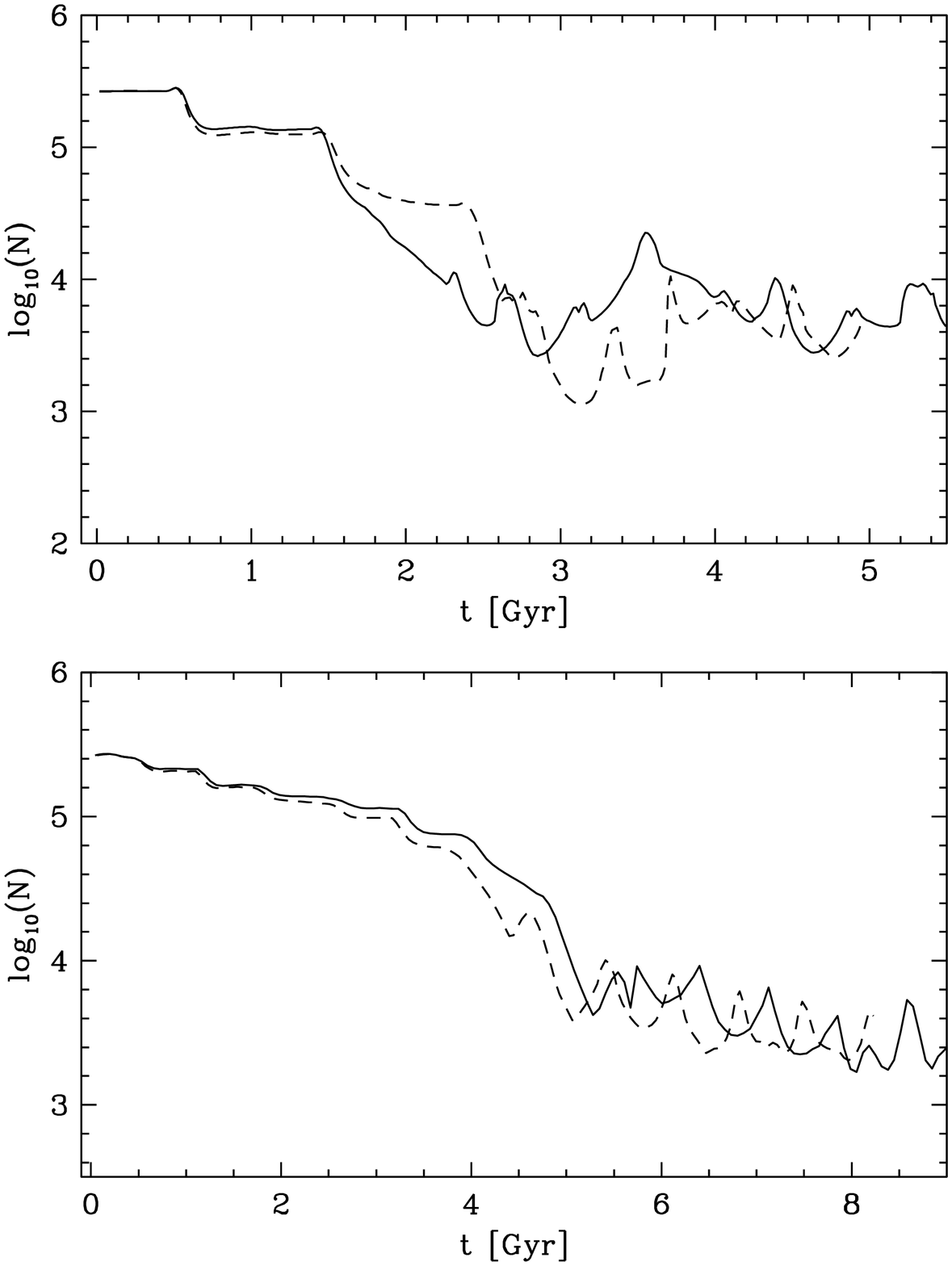}
\caption{\label{fig:N2}}
\end{figure}

{\normalsize F{\footnotesize IG}.}\ref{fig:N2} --- The number of
particles $N(t)$ in the volume with radius 0.8~kpc centred on the
density maximum of the satellite.  Upper panel: the solid line is
$32^3$~cell simulation RS1-1, and the dashed line is $64^3$~cell
simulation RS1-1L. Lower panel: the solid line is $32^3$~cell
simulation RS1-24, and the dashed line is $64^3$~cell simulation
RS1-24L. In both panels, the number of particles in RS1-1L and RS1-24L
is scaled down by the factor $(3\times10^5)/(2\times10^6)$, where the
nominator and denominator are the initial number of satellite
particles in the simulations with $32^3$ and $64^3$~cells,
respectively.

\begin{figure}[t]
\epsscale{0.4}
\plotone{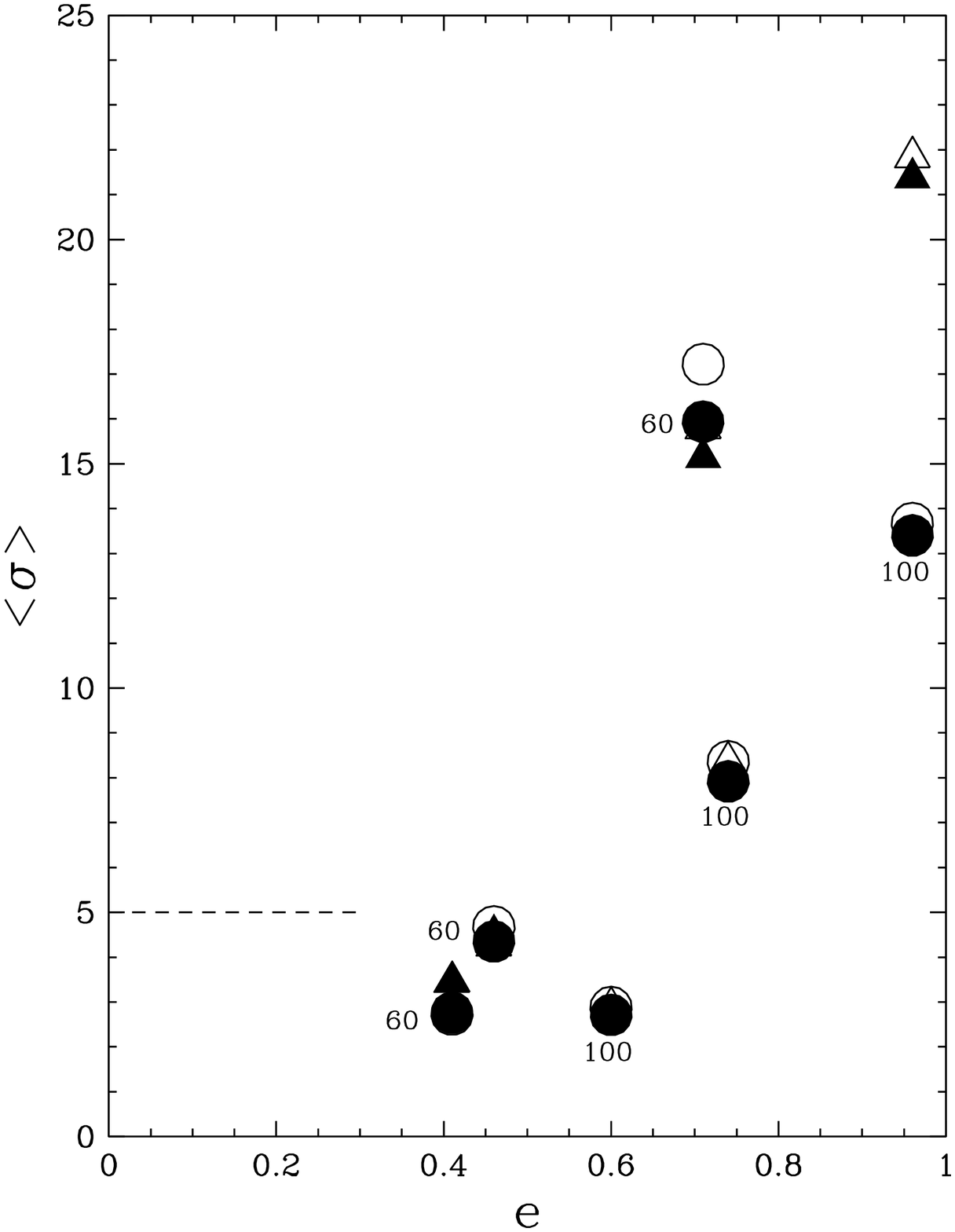}
\caption{\label{fig:vdisp_e}}
\end{figure}

{\normalsize F{\footnotesize IG}.}\ref{fig:vdisp_e} --- The
time-averaged velocity dispersion, computed over the first 2.5~Gyr
after $(M/L)_{\rm obs}\ge50$ is achieved, as a function of orbital
eccentricity, $e$. Solid circles are $\left< \sigma_0 \right>$ for all
{\sc Superbox} simulations presented here and in K97 with
$N_{\rm grid}=32^3$ cells and $N_{\rm
sat}=3\times10^5$~particles. Solid triangles: $\left<\sigma_0\right>$
for {\sc Grape} simulations ($e=0.46, 0.71$), and {\sc Superbox}
simulations with $N_{\rm grid}=64^3$ and $N_{\rm sat}=2\times10^6$
($e=0.41, 0.96$). Open circles show $\left<\sigma_{r_{1/2}}\right>$
for all {\sc Superbox} simulations with $N_{\rm grid}=32^3$ cells and
$N_{\rm sat}=3\times10^5$~particles, and open triangles are the
corresponding {\sc Grape} and {\sc Superbox} simulations with $N_{\rm
grid}=64^3$ and $N_{\rm sat}=2\times10^6$.  The horizontal dashed line
is the initial central line-of-sight velocity dispersion. Numbers next
to the filled circles are $R_{\rm apo}$ in~kpc.

\begin{figure}[b]
\epsscale{0.5}
\plotone{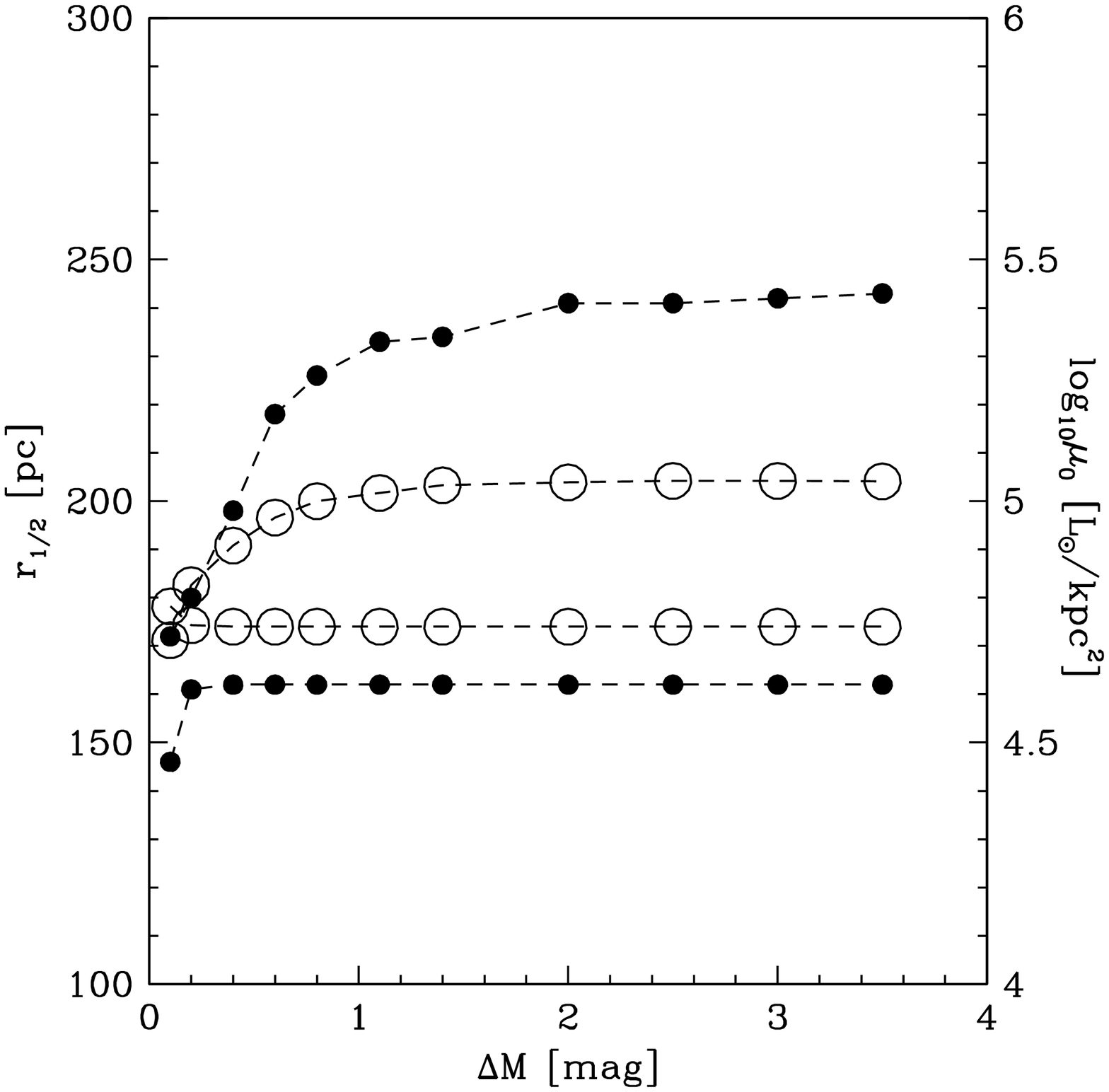}
\caption{\label{fig:delta-mag}}
\end{figure}

{\normalsize F{\footnotesize IG}.}\ref{fig:delta-mag} --- The
half-light radius, $r_{1/2}$ (filled circles), and central surface
brightness, $\mu_0$ (open circles), as a function of $\Delta M$.  The
set of two upper curves are for remnant RS1-4 at $t=6.27$~Gyr, and the
lower set of curves are for remnant RS1-5 at $t=8.74$~Gyr.  These data
are produced using $k=100$ bins within $r_{\rm bin}=4$~kpc and $(M/L)_{\rm
true}=3$.  This figure complements Fig.~12 in K97.

\begin{figure}[t]
\epsscale{0.4}
\plotone{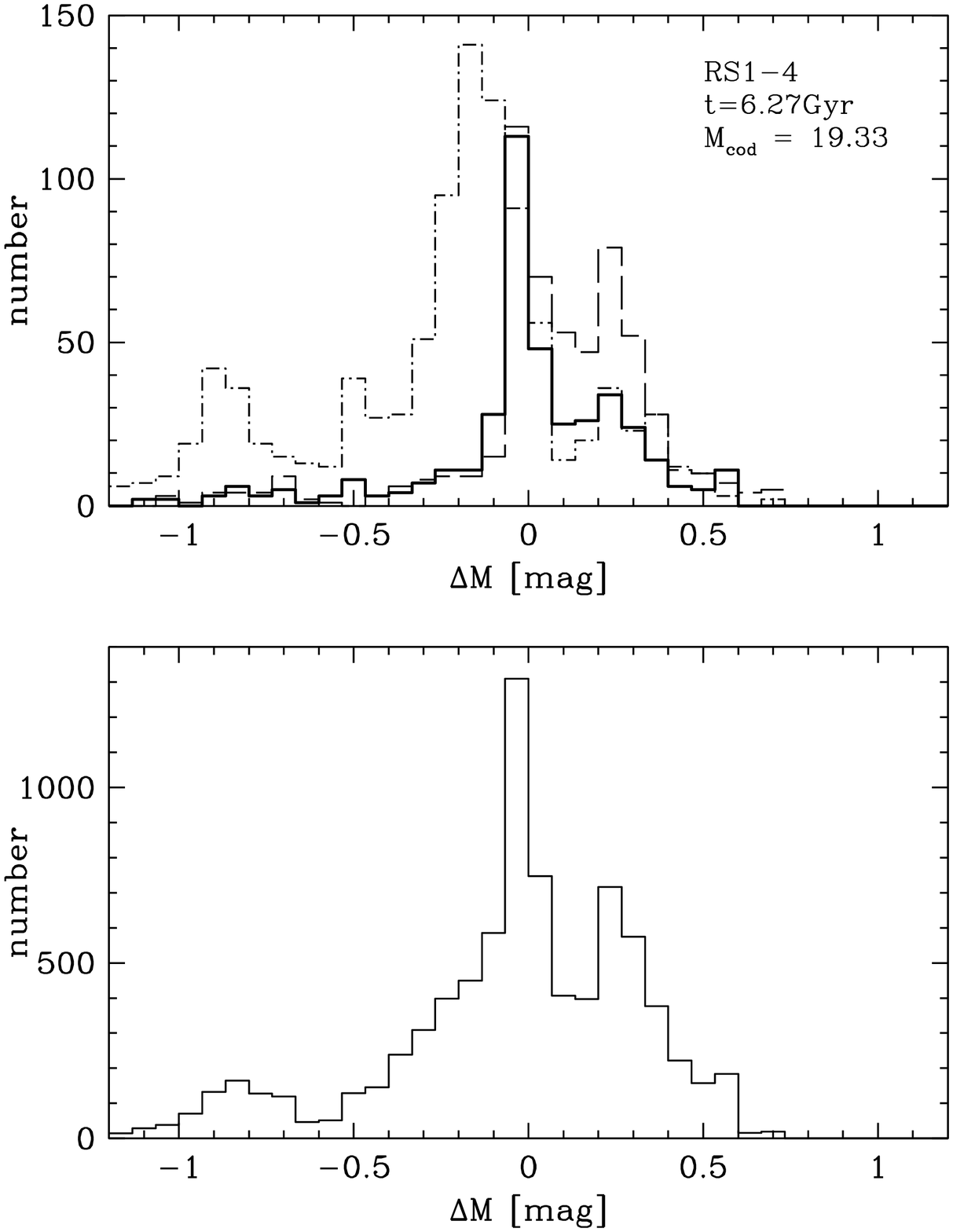}
\caption{\label{fig:number-dM-RS1-4}}
\end{figure}

{\normalsize F{\footnotesize IG}.}\ref{fig:number-dM-RS1-4} --- The
distance-modulus distribution of particles across the face of remnant
RS1-4. Both panels show the distribution of distance moduli relative
to the distance modulus of the remnant's density maximum. In the upper
panel, the solid histogram shows the distribution at the remnant's
centre, and the long-dashed and dot-dashed histograms are the
distributions in regions offset from the centre by 1.13~kpc along the
velocity gradient. The bottom panel shows the distribution for all
particles appearing projected within a radial distance of 1.2~kpc from
the position on the sky of the remnant's density maximum.  For details
see Section~\ref{subsubsec:hb}.  This figure complements figs.~9--12
in K97.

\begin{figure}[b]
\epsscale{0.4}
\plotone{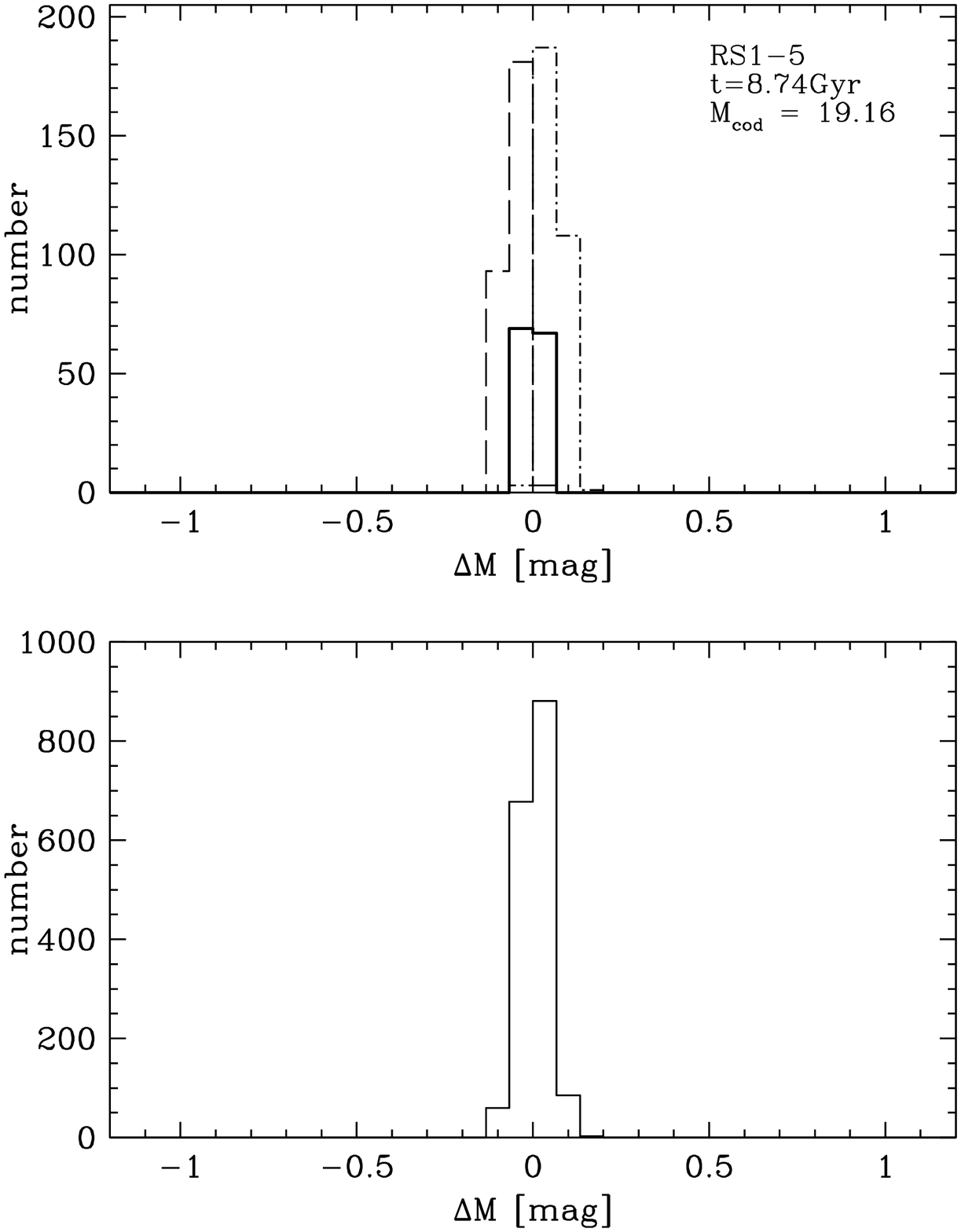}
\caption{\label{fig:number-dM-RS1-5}}
\end{figure}

{\normalsize F{\footnotesize IG}.}\ref{fig:number-dM-RS1-5} --- The
same as Fig.~\ref{fig:number-dM-RS1-4}, but for the snapshot of remnant
RS1-5.

\clearpage
\newpage

\clearpage

\end{document}